\documentclass[trackchanges, twocolumn]{aastex701}

\usepackage{siunitx}
\usepackage{amsmath}
\usepackage{gensymb}
\usepackage{rotating}
\usepackage[normalem]{ulem}
\begin{document}

\title{A Detailed Analysis of Intermediate-Scale Structure in Optical Extinction Curves: Expanded Census and Two-Family Classification}

\author[orcid=0000-0002-4634-5966,sname='Chamani M. Gunasekera']{Chamani M. Gunasekera}
\affiliation{Space Telescope Science Institute, 3700 San Martin Drive, Baltimore, MD 21218, USA}
\email[show]{cgunasekera@stsci.edu}  

\author[orcid=0000-0001-9462-5543, sname='Marjorie Decleir']{Marjorie Decleir} 
\affiliation{European Space Agency (ESA), ESA Office, Space Telescope Science Institute, 3700 San Martin Drive, Baltimore, MD 21218, USA}
\altaffiliation{ESA Research Fellow}
\email{mdecleir@stsci.edu}

\author[orcid=0000-0001-5340-6774,sname='Karl D. Gordon']{Karl D. Gordon}
\affiliation{Space Telescope Science Institute, 3700 San Martin Drive, Baltimore, MD 21218, USA}
\affiliation{Sterrenkundig Observatorium, Universiteit Gent, Krijgslaan 281 S9, B-9000 Gent, Belgium}
\email{kgordon@stsci.edu}

\author[0000-0002-0141-7436]{Geoffrey C. Clayton}
\affiliation{Space Science Institute,
4765 Walnut St., Suite B
Boulder, CO 80301, USA}
\email{
gclayton@spacescience.org}
\affiliation{Department of Physics \& Astronomy, Louisiana State University, Baton Rouge, LA 70803, USA}

\author[orcid=0000-0002-2449-0214,sname='Burcu G{\"u}nay']{Burcu G{\"u}nay}
\affiliation{Armagh Observatory and Planetarium, Armagh, BT61 9DB, UK}
\email{burcu.gunay@ege.edu.tr}

\author[orcid=0000-0002-5895-8268,sname='Dries Van De Putte']{Dries Van De Putte}
\affiliation{Department of Physics \& Astronomy, The University of Western Ontario, London ON N6A 3K7, Canada}
\affiliation{Institute for Earth and Space Exploration, The University of Western Ontario, London ON N6A 3K7, Canada}
\email{dvandepu@uwo.ca}

\author[orcid=0000-0002-9912-6046,sname='Petia Yanchulova']{Petia Yanchulova}
\affiliation{Institute of Astronomy, Bulgarian Academy of Sciences, 72, Tsarigradsko Chaussee Blvd, 1784, Sofia, Bulgaria}
\affiliation{University of Sofia, Faculty of Physics, 5 James Bourchier Blvd., 1164 Sofia, Bulgaria}
\email{petiay@gmail.com}

\begin{abstract}
The features of interstellar extinction curves serve as powerful diagnostics for interstellar dust, revealing information about its composition, size distribution, and the physical and chemical processes that shape it. 
%Extinction features are commonly classified based on their widths, as they probe different carriers. 
\citet{2020ApJ...891...67M} reported three faint but wide extinction features, termed intermediate-scale structures (ISS) at {4370, 4870, 6300\,\AA.} Since then, three additional ISS features have been reported in the literature at {7700, 5400, 8500\,\AA,} all with widths greater than any known diffuse interstellar band (DIB). 
We present new optical and UV Hubble Space Telescope/STIS spectra for a sample of 24 early-type OB stars. We used these data combined with 50 literature targets for a {systematic, homogeneous} analysis of ISS features, with the aim of investigating their observational behaviour to help constrain their carriers. This analysis revealed 9 more candidate ISS features.
We also find that ISS features can be arranged into two main families, which we call $\alpha$ and $\beta$, according to correlations between the feature strengths. Finally, we find strong correlations between the $2175$\,\AA\ bump strength and ISS features at {4353, 4847, 6443, 7710\,\AA,} suggesting their possible carbonaceous origin.
\end{abstract}


\section{Introduction}
\label{sec:iintro}
%\todo{tarball spectra + extinction, put in zenodo, zenodo link to the github, zenodo links in the paper. two different zenodo links for the data and github repo.}

Dust extinguishes and reddens starlight passing through nebulae, and plays a critical role in a multitude of astrophysical processing, including photoelectric heating, %extinction of starlight, 
collisional heating/cooling, charge exchange, formation of ${\rm H}_2$ and other  molecules, and molecular freeze-out \citep{2011piim.book.....D}.
%These processes depend on various interstellar dust properties, such as the grain composition and size distribution,
%in turn impacting our observations and influencing our understanding of the chemical evolution of the cosmos. 
Dust extinction being a fundamental probe of dust grain composition, size, and shape, provides essential information to correct for these extinguishing and reddening effects, and powerful insight to the chemical composition and thermal balance of the interstellar medium (ISM).
%Extinction measurements from far-UV through mid-IR wavelengths illustrate grains extinguish light of all wavelengths to varying efficiencies \citep{2023ApJ...950...86G}.
%Dust extinction includes both absorption and scattering of starlight out of the line-of-sight. 
Extinction curves include broad features caused by different types of dust grains. %, as well as absorption features from individual atoms and molecules. 
Extinction features tend to be categorized based on width, {defined here as the full width at half maximum (FWHM)}, as they probe different carriers \citep{2001ApJ...548..296W,2001ApJ...554..778L,2025arXiv250707162S}. 
Unlike gas-phase atoms or small molecules which have discrete electronic transitions resulting in very narrow features, larger molecules such as polycyclic aromatic hydrocarbons (PAHs) or fullerenes tend to produce broader features, while solid-state dust particles have even broader bands.
Extinction features at optical wavelengths include %narrow interstellar absorption features from Na, Ca and K, and 
diffuse interstellar bands (DIBs) \citep{1995ARA&A..33...19H, 2025arXiv250707162S}, with widths on the scale of {$\sim 10-100$\,\AA.}
Also in the optical, a depression of approximate width {$> 1000$\,\AA,} termed the Very Broad Structure (VBS) is located {at $5000\lesssim\lambda\lesssim6700$\,\AA\, as} noted in both \citet{1966ApJ...144..305W} and \citet{2020ApJ...891...67M}, with no confirmed origin. 
On the intermediate scale, the well-known $2175$\,\AA\ with a width of $\sim 427$\,\AA, is thought to come from carbonaceous grains \citep{1965ApJ...142.1681S, 2001ApJ...554..778L, 2011ApJ...742....2S}.

Three intermediate scale structures (ISS) at {4370, 4870, 6300\,\AA\ with widths 464, 424, 964\,\AA\ respectively,} were analysed and reported by \citet[hereafter M20]{2020ApJ...891...67M}. Their analysis found that the strength of two of the three of these ISS features correlates with the $2175$\,\AA\ bump strength, but none were correlated with the total-to-selective extinction ratio, $R(V)$. 
\citet[hereafter M21]{2021MNRAS.501.2487M} identified a fourth extinction feature that was wider than any known DIB, %but narrower than the broadband variability, 
at {7700\,\AA, with a width of $\sim 177$\,\AA.} This feature was also noted in \citet[hereafter F19]{2019ApJ...886..108F}, but was labelled a DIB at the time. This study finds that the {7700\,\AA\ }ISS feature is faint or absent in molecular carbon-rich environments.
\citet[hereafter Z24]{2024ApJ...971..127Z} independently confirm the {7700\,\AA\ }feature and identify an additional feature at {5400\,\AA,} using Gaia XP Spectra{, for which they do not specify a width, but do speculate that it is an ISS feature}.
A sixth ISS feature, at {8500\,\AA} {with width of $\sim 490$\,\AA} was noted in \citet[hereafter G25]{2025ApJ...988....5G} using Gaia low-resolution XP spectra. These latter two features remain to be confirmed in high-resolution spectra.
%We present a summary of the above described known ISS features in Table~\ref{tab:iss}. 

The primary goal of the present paper is to conduct a {systematic, homogeneous} analysis of the six literature ISS features, as well as additional candidate ISS features identified in this work. {As far as we are aware, this paper will constitute only the second study, following M20, to solely focus on ISS features.}
We aim to characterise the observational behaviour of ISS features, and to confirm the new literature features at {5400\,\AA\ }and  {8500\,\AA}, along with the candidate features.
This paper is organised as follows. We use a combination of low-resolution \textit{Hubble Space Telescope/Space Telescope Imaging Spectrograph} (HST/STIS) and \textit{International Ultraviolet Explorer} (IUE) spectrophotometry, for a sample of 74 lines of sight. Section~\ref{sec:data} discusses this selected sample, and the required data reduction procedures. To detect the ISS features, we measured and fitted the extinction curves of all 74 sightlines as described in Section~\ref{sec:ext_curve}. We then analyse the ISS features in Section~\ref{sec:analysis}. Finally, in Section~\ref{sec:summary}, we present a summary of our findings on the nature of the 6 literature ISS features and the newly detected candidate ISS features.

\section{Data}
\label{sec:data}

\begin{table*}
\centering
\caption{The \textit{Prince} sample of stars. \label{tab:prince}}
\begin{tabular}{lcccccc}
\hline
\hline
Star & Spectral Type & ${\rm V}$ (mag) & ${\rm E(B-V)}$ & RA & Dec & Reference\footnote{\textbf{References}--(1)~\citet{2004ApJ...616..912V}; (2)~\citet{2021ApJ...916...33G}; (3)~\citet{2022ApJ...930...15D}; (4)~\citet{1996ApJ...472..755L}; (5)~\citet{1990ApJS...72..163F}} \\
\hline
ALS882    & B2V   &   11.82  &  0.46  &  07 57 3.4967    &  -27 54 44.27 &    \\
ALS1795   & B1.5V &   10.16  &  0.53  &  10 42 50.1555   &  -59 25 31.11 &    \\
ALS2285   & B2V   &   10.22  &  0.75  &  11 16 44.3959   &  -59 35 16.92 &    \\
ALS6028   & B0V   &   11.06  &  0.77  &  00 05 0.8472    &  +63 49 33.19 &    \\
ALS6206   & O9.5V &   11.04  &  1.32  &  00 27 16.9991   &  +64 42 19.64 &    \\
ALS6213   & B1V   &   11.11  &  1.24  &  00 28 11.1464   &  +64 07 51.81 &    \\
ALS6672   & B1V   &   10.02  &  0.34  &  01 39 39.4728   &  +57 49 47.25 &    \\
ALS8351   & B2V   &   10.98  &  0.38  &  05 27 16.7828   &  +34 30 56.80 &    \\
ALS13253  & B1V   &   11.8   &  0.42  &  23 52 45.1006   &  +61 09 52.58 &    \\
ALS15273  & B2V   &   11.01  &  0.83  &  20 19 14.7126   &  +39 09 0.60  &    \\
ALS18106  & B2V   &   11.92  &  0.29  &  09 16 56.1408   &  -62 26 22.90 &    \\
ALS18098  & B1.5V &   11.77  &  0.37  &  23 52 42.0139   &  +61 06 17.27 &    \\
BD+56 510 & B5V   &   9.32   &  0.47  &  02 18 47.7725   &  +57 08 6.73  &  1 \\
HD036982  & B1.5V &   8.46   &  0.38  &  05 35 9.8381    &  -05 27 53.19 &  1 \\
HD037021  & B0V   &   7.96   &  0.54  &  05 35 16.1353   &  -05 23 6.76  &  1 \\
HD038087  & B3II  &   8.3    &  0.3   &  05 43 00.5755   &  -02 18 45.39 &  5 \\
HD093160  & O7V   &   7.88   &  0.33  &  10 44 07.2603   &  -59 34 30.62 &  1 \\
HD111934  & B2Ib  &   6.86   &  0.49  &  12 53 37.6095   &  -60 21 25.44 &  1 \\
HD192660  & B0Ia  &   7.38   &  0.87  &  20 14 26.0772   &  +40 19 44.96 &  2 \\
HD204827  & B0V   &   7.95   &  1.00  &  21 28 57.7592   &  +58 44 23.24 &  3 \\
HD210121  & B3V   &   7.67   &  0.38  &  22 08 11.9028   &  -03 31 52.77 &  4 \\
HD239689  & B5V   &   8.84   &  0.34  &  21 30 45.9265   &  +57 12 00.16 &  1 \\
HD294264  & B3V   &   9.47   &  0.5   &  05 35 13.3447   &  -04 51 44.92 &  5 \\
WALKER67  & B2V   &   10.8   &  0.85  &  06 40 37.2542   &  +09 47 29.77 &  1 \\
\hline
\end{tabular}
\end{table*}
%%%%%%%%%%%%%%%%%

\begin{figure*}
  \centering
  \includegraphics[width=1.0\textwidth]{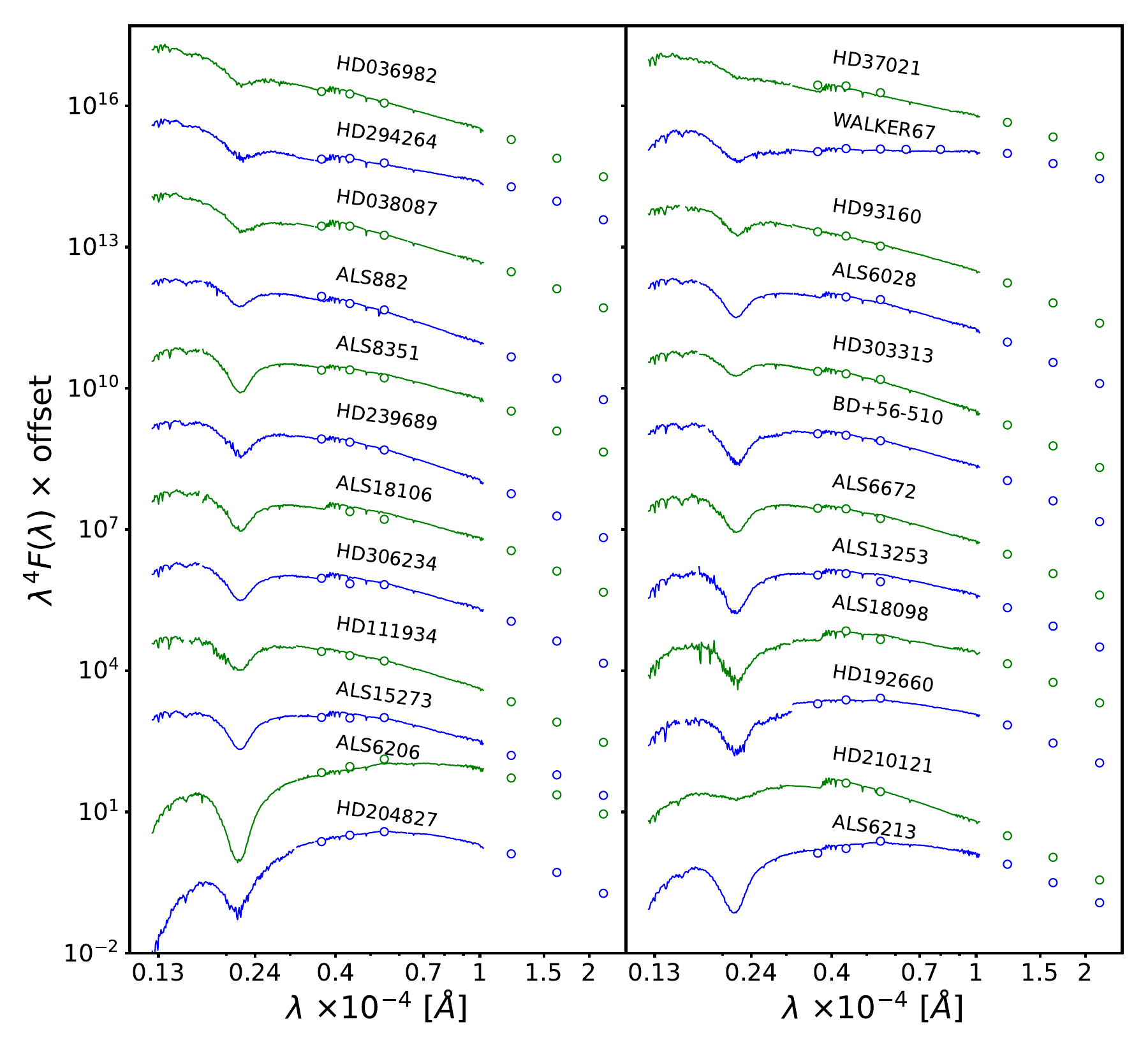}
      \caption{Optical and UV spectra of the {24} stars in the \textit{Prince} sample. The HST/STIS spectra are shown as solid lines, and the photometric data (U, B, V, J, H, K) are shown as open circles. The blue and green colors are simply used to distinguish between adjacent curves, and do not carry any particular meaning. The spectra are roughly ordered according to the size of the UV bump.}
    \label{fig:stacked_spec}
\end{figure*}

\subsection{Sample Selection}

We use 50 out of the 72 lines of sight in the  F19 sample, omitting the stars with effective temperatures less than 20,000~K, focusing on stars well modelled with non-LTE stellar model atmospheres (further context is provided in Section~\ref{sec:measuring_ext}). We extend this sample with an additional 24 lines of sight with low-resolution HST/STIS spectroscopic data in  both UV and optical (hereafter, ``\textit{Prince}''\footnote{During a dispute with Warner Bros, the musician Prince, changed his name to his now very famous unpronounceable glyph known as the Prince symbol (see \url{https://en.wikipedia.org/wiki/Prince_(musician)}). We borrow this terminology to call our additional sample the \textit{Prince} sample.} sample). Figure~\ref{fig:stacked_spec} shows the optical and UV spectra for the full \textit{Prince} sample as solid lines, including the available Johnson photometry data (U, B, V, J, H, K) as open circles. This sample, PID 17516 (PI: Decleir), was originally selected to study sightlines with very high and low total-to-selective extinction ${\rm R(V)}$. However, the sample selection method preferentially found stars with large V-band uncertainties, which artificially inflated or suppressed ${\rm R(V)}$. Since the data is of good quality we use it for analysing ISS features instead.
Thus, the sample was renamed from the high/low $R(V)$ sample to the Prince sample.
For four overlapping targets in the two samples, we preferred the newly obtained STIS data to derive the extinction curve as these have higher-resolution and S/N in the UV regime, relative to the IUE data available for the remainder of the F19 sample. 

The \textit{Prince} sample was carefully chosen from existing literature samples
\citep{1990ApJS...72..163F, 1996ApJ...472..755L, 2004ApJ...616..912V, 2021ApJ...916...33G, 2022ApJ...930...15D} to probe high- and low-${\rm R(V)}$ sightlines. 
Six sightlines were chosen based on ${\rm R(V)}$ measurements reported in: \citet{2022ApJ...930...15D, 2021ApJ...916...33G, 1990ApJS...72..163F, 2009ApJ...705.1320G}, in order of priority. 
An additional nine sightlines were added from \citet{2004ApJ...616..912V} after re-measuring the extinction curves and ${\rm R(V)}$ values. The lowest known ${\rm R(V)}$ sightline HD 210121
\citep{1996ApJ...472..755L} was added bringing the total to 16 lines-of-sight. Thirteen more from the second installment of the ALMA Luminous Star (ALS) catalogue \citep{2021MNRAS.504.2968P} were added to the sample. The catalogue stars were whittled down by selecting targets with spectral types of O9V to B3V, which are UV and optically bright with well-modelled spectral shapes, and those with sufficiently high dust extinction (i.e. ${\rm E(B-V)} > 0.3$) ensuring reliable extinction measurements. 
Also omitted from the catalogue were stars appearing to have an inconsistency between their spectral types and spectra, as well as those showing emission lines in their Gaia XP spectra, emission-line stars, and X-ray or eclipsing binaries. Of the 29 selected targets, five observations were unsuccessful, leaving a total of 24 new optical and UV STIS spectra for analysis. Table~\ref{tab:prince} presents the properties of these 24 targets in the \textit{Prince} sample.

%* from G23: \\
%    - 2 LOS w/ R(55) <= 2.5\\
%    - 4 LOS w/R(55) >= 5\\
%* Add from Valencic 2004 \\
%    - 3 LOS w/ R(55) <= 2.5\\
%    - 6 LOS w/R(55) >= 5\\
%* lowest known RV: HD210121, Larson et al., 1996\\
%* from OB star catalog from Pantaleoni Gonza lez et al. (2021)\\
%    - select O9V to $B_3$V, bright at UV-optical + simple spectral shapes compared to later type stars
%    - select E(B-V)> 0.3, stars with sufficiently high dust extinction to ensure reliable extinction measurements\\
%    - visual comparison between the Gaia BPRP spectra (De Angeli et al., 2022) and the models,\\
%        - exclude stars from sample with spectra that seem inconsistent with their spectral types or with extinctions that are very far off as expected from their R(V ) values.\\
%        - excluded stars that show emission lines in their BPRP spectra, as well as stars that were flagged as emission-line star, x-ray binary or eclipsing binary on SIMBAD,\\
%    - 8 LOS w/ R(55) <= 2.5\\
%    - 5 LOS w/R(55) >= 5\\

\subsection{Wavelength Coverage and Instrumentation}
We obtained G140L and G230L STIS/MAMA data covering the spectral range $1150$--$1730$\,\AA\ and $1570$--$3180$\,\AA\ at resolving power $960$--$1440$\ and $500$--$1010$\ respectively, and G430L and G750L STIS/CCD data covering the range $2900$--$5700$\,\AA\ and $5240$--$10270$\,\AA\ both of which have resolving power $530$--$1040$. This study requires sufficient S/N to allow in-depth examination of the faint extinction features, so the observations were designed to achieve a %. Thus, with the goal of a 
S/N of $10$ at a resolution of $100$ using the above gratings.%, the UV spectra were obtained using the G140L/FUV MAMA, G230L/NUV MAMA or G230LB/CCD gratings. 
This resolution and S/N was chosen as it is sufficient to measure extinction features such as the UV bump strength, width and wavelength. 
Likewise, aiming for a S/N of $100$ at a resolution of $100$, the optical spectra were obtained using G430L/CCD and G750L/CCD.

\subsection{Fringe Correction}
Prior to any fits, the $24$ targets in the \textit{Prince} sample required fringe corrections in the long wavelength ends of the stellar spectra. The fringing particularly affects the Paschen series, worsening the stellar spectral fitting. The standard STIS defringing pipeline \citep{stisdhb2024} was used to mitigate this instrumental anomaly.

For the F19 sample, we obtained the data as reduced by F19 that included STIS defringing. The processed data for the \textit{Prince} sample can be obtained from Zenodo \citep{gunasekera2026prince}, and the raw data from MAST archive at \dataset[doi:10.17909/cj0j-eh94]{https://dx.doi.org/10.17909/cj0j-eh94}.

\section{Method}
\label{sec:ext_curve}

Historically, extinction curves have been derived using the ``pair method'' \citep{1983ApJ...266..662M,1992AJ....104.1916C}, by comparing the spectral energy distribution (SED) of a reddened star to that of an unreddened (or mildly reddened) one with the same spectral type. To overcome the difficulty of finding good spectral matches \citet{2005AJ....130.1127F} introduced a method to use stellar atmosphere models as the comparison star. {Both \citet{2005AJ....130.1127F} and \citet{2019ApJ...886..108F} show that using model atmospheres in-place of lightly reddened stars produces sufficiently accurate extinction curves.} This method begins with forward modelling of the stellar spectrophotometry to determine the best fit model atmosphere. 
For the present sample of stars we use the latest TLUSTY non-LTE stellar model atmospheres \citep{2025AJ....169..178H}. These model atmospheres are then fit using v1.4 of the \texttt{measure\_extinction} package \citep{karl_gordon_2025_15831932}. This package uses the $R(V)$--dependent extinction curve across all wavelengths from \citet{2023ApJ...950...86G}, which is based on earlier studies \citep{2009ApJ...705.1320G, 2019ApJ...886..108F, 2021ApJ...916...33G, 2022ApJ...930...15D}.

\subsection{Determining the Stellar Parameters}
\label{sec:measuring_ext}
For this work, we fit the stellar SED in the optical for all 74 stars, constraining the effective temperature (${T_\mathrm{eff}}$), surface gravity ($\log(g)$), the total extinction in the V band ($A(V)$), the microturbulent velocity ($v_\mathrm{turb}$), and the macroscopic velocity field (Velocity). We use the stellar type to estimate Gaussian priors for $\log{T_\mathrm{eff}}$ and $\log(g)$, with $\log{T_\mathrm{eff}}$ truncated at $\pm3\sigma$ about its prior value.
We keep the remainder of the stellar atmosphere model parameters fixed including the metallicity at Solar since our sample consists of stars in the Milky Way. The fitting was done in two steps. First, preliminary best-fit parameters for the above listed stellar model parameters
were obtained using the damped least-squares (DLS) fitting method.\footnote{{DLS fitting was done with \texttt{LevMarLSQFitter} from Astropy \citep{2022ApJ...935..167A}}} These parameters were then inputs for the Markov
Chain Monte Carlo (MCMC) fitting\footnote{{MCMC fitting was conducted with the \texttt{emcee} python package \citep{2013PASP..125..306F}}}. We used $10^5$ steps, and the \texttt{measure\_extinction} package defaults of a burn fraction of 50$\%$, and the number of walkers equal to twice the number of parameters in the model (which in this case is $2\times6$). 
Although the $10^5$ steps did not satisfy the 50 times the autocorrelation time criterion in \cite{2013PASP..125..306F}, extending the chain length to $6\times10^5$ steps (which increases the run-time by a factor of 10) only changed the fitted parameter uncertainties by $\sim20$\% on average. The trace plots also show that the chains have relaxed, even at $10^5$ steps, so we determined this chain length provides adequate convergence.

\startlongtable
\begin{deluxetable*}{lcccc}
\tablecaption{Best-fit Stellar Parameters.
\label{tab:star_fit}}
\tablehead{
\colhead{Starname} &
\colhead{$\log T_{\mathrm{eff}}$} &
\colhead{$\log g$} &
\colhead{$v_{\mathrm{turb}}$} &
\colhead{Velocity} \\
\colhead{} &
\colhead{} &
\colhead{} &
\colhead{(km s$^{-1}$)} &
\colhead{(km s$^{-1}$)}
}
\startdata
ALS882 & 4.362$\pm$0.002 & 4.149$\pm$0.020 & 2.691$\pm$0.526 & 9.933$\pm$0.965 \\
ALS1795 & 4.386$\pm$0.005 & 4.149$\pm$0.017 & 3.193$\pm$0.352 & 7.019$\pm$0.932 \\
ALS2285 & 4.304$\pm$0.002 & 3.730$\pm$0.036 & 2.861$\pm$0.667 & 2.236$\pm$0.968 \\
ALS6028 & 4.320$\pm$0.031 & 4.059$\pm$0.056 & 3.936$\pm$0.857 & 5.305$\pm$1.017 \\
ALS6206 & 4.539$\pm$0.005 & 4.429$\pm$0.053 & 2.223$\pm$0.663 & -0.459$\pm$0.472 \\
ALS6213 & 4.389$\pm$0.051 & 4.267$\pm$0.039 & 4.545$\pm$2.075 & 2.435$\pm$0.987 \\
ALS6672 & 4.360$\pm$0.007 & 3.589$\pm$0.018 & 3.370$\pm$0.453 & 3.582$\pm$0.968 \\
ALS8351 & 4.446$\pm$0.002 & 4.142$\pm$0.023 & 2.785$\pm$0.484 & 4.000$\pm$1.032 \\
ALS13253 & 4.311$\pm$0.016 & 3.944$\pm$0.029 & 4.061$\pm$0.180 & 2.659$\pm$1.916 \\
ALS15273 & 4.179$\pm$0.002 & 4.453$\pm$0.009 & 5.877$\pm$0.025 & 5.505$\pm$1.100 \\
ALS18106 & 4.313$\pm$0.006 & 3.951$\pm$0.039 & 3.334$\pm$0.641 & 5.867$\pm$1.006 \\
ALS18098 & 4.202$\pm$0.006 & 3.995$\pm$0.046 & 4.597$\pm$0.114 & -3.716$\pm$0.976 \\
BD+56-510 & 4.279$\pm$0.014 & 4.116$\pm$0.009 & 7.474$\pm$1.253 & 3.019$\pm$1.244 \\
HD036982 & 4.323$\pm$0.004 & 4.186$\pm$0.010 & 3.785$\pm$0.187 & 265.720$\pm$5.157 \\
HD037021 & 4.226$\pm$0.003 & 4.198$\pm$0.019 & 2.331$\pm$0.303 & 174.024$\pm$4.664 \\
HD038087 & 4.203$\pm$0.001 & 4.356$\pm$0.007 & 3.351$\pm$0.430 & -23.770$\pm$2.483 \\
HD093160 & 4.685$\pm$0.007 & 4.072$\pm$0.075 & 2.785$\pm$0.482 & 1.520$\pm$1.086 \\
HD111934 & 4.381$\pm$0.005 & 2.996$\pm$0.079 & 7.579$\pm$1.782 & 2.490$\pm$1.013 \\
HD192660 & 4.443$\pm$0.006 & 2.790$\pm$0.060 & 6.249$\pm$1.312 & 2.517$\pm$0.941 \\
HD204827 & 4.342$\pm$0.020 & 3.729$\pm$0.070 & 5.327$\pm$1.319 & -1.462$\pm$1.176 \\
HD210121 & 4.177$\pm$0.001 & 3.983$\pm$0.007 & 2.024$\pm$0.019 & -5.288$\pm$0.953 \\
HD239689 & 4.334$\pm$0.024 & 4.133$\pm$0.030 & 3.946$\pm$0.230 & 5.284$\pm$0.983 \\
HD294264 & 4.273$\pm$0.001 & 4.438$\pm$0.005 & 3.568$\pm$0.357 & -54.780$\pm$4.001 \\
WALKER67 & 4.203$\pm$0.066 & 4.671$\pm$0.079 & 5.634$\pm$0.937 & 0.000$\pm$0.001 \\
BD+56d517 & 4.365$\pm$0.003 & 3.829$\pm$0.027 & 2.848$\pm$0.689 & -2.848$\pm$0.905 \\
BD+56d518 & 4.336$\pm$0.003 & 3.844$\pm$0.009 & 3.009$\pm$0.615 & -0.884$\pm$0.725 \\
BD+56d576 & 4.355$\pm$0.001 & 3.502$\pm$0.018 & 2.811$\pm$0.540 & -2.357$\pm$0.930 \\
CPD-41d7715 & 4.374$\pm$0.001 & 4.025$\pm$0.004 & 2.014$\pm$0.012 & -6.001$\pm$0.965 \\
CPD-57d3507 & 4.311$\pm$0.009 & 4.107$\pm$0.013 & 7.479$\pm$0.017 & -4.062$\pm$1.106 \\
CPD-57d3523 & 4.361$\pm$0.018 & 3.527$\pm$0.060 & 7.966$\pm$2.813 & -4.228$\pm$0.960 \\
CPD-59d2591 & 4.606$\pm$0.006 & 4.462$\pm$0.041 & 3.071$\pm$0.228 & -0.770$\pm$1.096 \\
CPD-59d2600 & 4.711$\pm$0.004 & 4.106$\pm$0.017 & 2.684$\pm$0.451 & -1.603$\pm$0.946 \\
CPD-59d2625 & 4.359$\pm$0.002 & 4.370$\pm$0.008 & 3.283$\pm$0.405 & -3.127$\pm$0.975 \\
GSC03712-01870 & 4.524$\pm$0.002 & 4.156$\pm$0.026 & 2.831$\pm$0.494 & -0.195$\pm$0.919 \\
HD013338 & 4.354$\pm$0.001 & 3.646$\pm$0.005 & 2.457$\pm$0.480 & -7.395$\pm$0.930 \\
HD014250 & 4.377$\pm$0.003 & 3.491$\pm$0.045 & 2.596$\pm$0.588 & -7.270$\pm$0.924 \\
HD014321 & 4.355$\pm$0.002 & 3.621$\pm$0.009 & 2.599$\pm$0.488 & -3.937$\pm$0.963 \\
HD027778 & 4.178$\pm$0.004 & 4.025$\pm$0.059 & 4.597$\pm$1.051 & -4.715$\pm$0.876 \\
HD028475 & 4.176$\pm$0.001 & 4.250$\pm$0.001 & 2.000$\pm$0.001 & -0.540$\pm$0.563 \\
HD030122 & 4.177$\pm$0.001 & 3.722$\pm$0.005 & 2.039$\pm$0.029 & -2.451$\pm$0.937 \\
HD030675 & 4.178$\pm$0.002 & 4.015$\pm$0.008 & 2.212$\pm$0.127 & -0.258$\pm$1.968 \\
HD037061 & 4.496$\pm$0.001 & 4.560$\pm$0.041 & 2.601$\pm$0.597 & -1.330$\pm$0.877 \\
HD040893 & 4.444$\pm$0.006 & 3.413$\pm$0.021 & 8.339$\pm$0.564 & -1.751$\pm$0.925 \\
HD046106 & 4.467$\pm$0.034 & 4.145$\pm$0.033 & 3.378$\pm$0.519 & 1.060$\pm$2.339 \\
HD046660 & 4.311$\pm$0.015 & 4.067$\pm$0.026 & 5.418$\pm$0.638 & -1.285$\pm$1.690 \\
HD054439 & 4.383$\pm$0.003 & 3.850$\pm$0.027 & 2.645$\pm$0.486 & -4.465$\pm$0.978 \\
HD062542 & 4.186$\pm$0.006 & 4.133$\pm$0.031 & 3.070$\pm$0.927 & -2.195$\pm$0.948 \\
HD068633 & 4.263$\pm$0.002 & 3.922$\pm$0.007 & 3.989$\pm$0.305 & 3.243$\pm$0.926 \\
HD070614 & 4.248$\pm$0.001 & 3.925$\pm$0.003 & 4.628$\pm$0.007 & -4.202$\pm$1.381 \\
HD091983 & 4.373$\pm$0.002 & 3.501$\pm$0.045 & 2.154$\pm$0.407 & -0.805$\pm$0.718 \\
HD092044 & 4.367$\pm$0.006 & 3.508$\pm$0.018 & 7.537$\pm$0.711 & -6.388$\pm$0.930 \\
HD093028 & 4.533$\pm$0.004 & 4.176$\pm$0.031 & 2.893$\pm$0.460 & -1.907$\pm$1.383 \\
HD093222 & 4.546$\pm$0.004 & 3.780$\pm$0.080 & 2.991$\pm$0.313 & -0.844$\pm$0.849 \\
HD104705 & 4.439$\pm$0.001 & 3.500$\pm$0.001 & 9.999$\pm$0.002 & -2.586$\pm$1.168 \\
HD110946 & 4.255$\pm$0.001 & 3.000$\pm$0.003 & 9.995$\pm$0.004 & -5.168$\pm$0.945 \\
HD142096 & 4.241$\pm$0.001 & 4.383$\pm$0.005 & 3.728$\pm$0.729 & 2.437$\pm$1.085 \\
HD147889 & 4.211$\pm$0.023 & 4.635$\pm$0.061 & 5.714$\pm$0.316 & -0.226$\pm$0.258 \\
HD149452 & 4.583$\pm$0.002 & 3.899$\pm$0.151 & 7.067$\pm$2.117 & -2.519$\pm$1.112 \\
HD164073 & 4.240$\pm$0.001 & 4.125$\pm$0.001 & 2.050$\pm$0.049 & 0.437$\pm$2.456 \\
HD172140 & 4.449$\pm$0.004 & 4.000$\pm$0.018 & 8.495$\pm$0.614 & -4.514$\pm$1.011 \\
HD193322 & 4.455$\pm$0.005 & 3.805$\pm$0.030 & 6.244$\pm$1.617 & -4.158$\pm$1.209 \\
HD197512 & 4.380$\pm$0.003 & 4.000$\pm$0.012 & 2.000$\pm$0.007 & -2.551$\pm$0.977 \\
HD197702 & 4.305$\pm$0.001 & 2.624$\pm$0.004 & 7.826$\pm$0.364 & -8.642$\pm$1.032 \\
HD198781 & 4.439$\pm$0.024 & 3.750$\pm$0.043 & 10.000$\pm$0.007 & -4.288$\pm$6.138 \\
HD199216 & 4.350$\pm$0.007 & 3.207$\pm$0.118 & 8.084$\pm$2.087 & -3.020$\pm$1.772 \\
HD210072 & 4.182$\pm$0.001 & 3.653$\pm$0.026 & 6.625$\pm$0.668 & -6.237$\pm$1.031 \\
HD217086 & 4.609$\pm$0.008 & 4.152$\pm$0.154 & 7.391$\pm$0.306 & 1.957$\pm$2.371 \\
HD220057 & 4.282$\pm$0.001 & 4.333$\pm$0.005 & 2.110$\pm$0.147 & -5.883$\pm$0.911 \\
HD228969 & 4.371$\pm$0.005 & 3.517$\pm$0.048 & 2.519$\pm$0.752 & 2.359$\pm$2.950 \\
HD236960 & 4.439$\pm$0.001 & 3.750$\pm$0.001 & 9.999$\pm$0.001 & -3.249$\pm$1.104 \\
HD239693 & 4.283$\pm$0.004 & 4.232$\pm$0.036 & 2.010$\pm$0.016 & -2.500$\pm$1.058 \\
HD239722 & 4.293$\pm$0.005 & 3.769$\pm$0.025 & 2.057$\pm$0.114 & -0.918$\pm$0.631 \\
HD239745 & 4.370$\pm$0.021 & 4.221$\pm$0.031 & 7.398$\pm$1.041 & -2.392$\pm$1.014 \\
HD292167 & 4.553$\pm$0.005 & 3.053$\pm$0.049 & 7.408$\pm$0.028 & -1.753$\pm$1.065 \\
HD303068 & 4.377$\pm$0.002 & 4.024$\pm$0.040 & 2.236$\pm$0.300 & 0.714$\pm$1.283 \\
NGC2244-VS11 & 4.440$\pm$0.009 & 4.313$\pm$0.027 & 3.503$\pm$0.088 & -3.496$\pm$0.976 \\
TRUMPLER14-6 & 4.439$\pm$0.001 & 4.001$\pm$0.001 & 9.998$\pm$0.002 & -0.405$\pm$0.889 \\
TRUMPLER14-27 & 4.404$\pm$0.010 & 4.436$\pm$0.033 & 3.242$\pm$0.444 & -1.432$\pm$0.803 \\
\enddata
\end{deluxetable*}

The effects of varying $\log{T_{\mathrm{eff}}}$ and $\log(g)$ are most prominently observed in the Balmer series of the stellar spectrum. In contrast, $A(V)$ primarily influences the overall slope of the stellar SED and therefore has a stronger impact on the J, H, and K band photometry. To account for these differences, we apply appropriate weights during the stellar spectra fitting process. By default, \texttt{measure\_extinction} computes the uncertainty-weighted fits, with weights proportional to $1/(\sigma_{F_\lambda})^2$%$1/(\delta F_\lambda)^2$
. For this work, we modify the default by weighting the photometry five times more heavily to balance the influence of the more numerous spectroscopic points, and to counter the lack of covariance in fitting the spectra (i.e. the fitting procedure treats each spectral point as independent even though neighbouring points may by partially correlated).
We also add a 1$\%$ uncertainty to the STIS spectra, which is the maximum S/N based on absolute flux considerations \citep{2014PASP..126..711B}.
{The best-fit stellar atmosphere model parameters for each star are presented in Table\,\ref{tab:star_fit}.} 
\begin{figure*}
  \centering
  \includegraphics[width=0.7\textwidth]{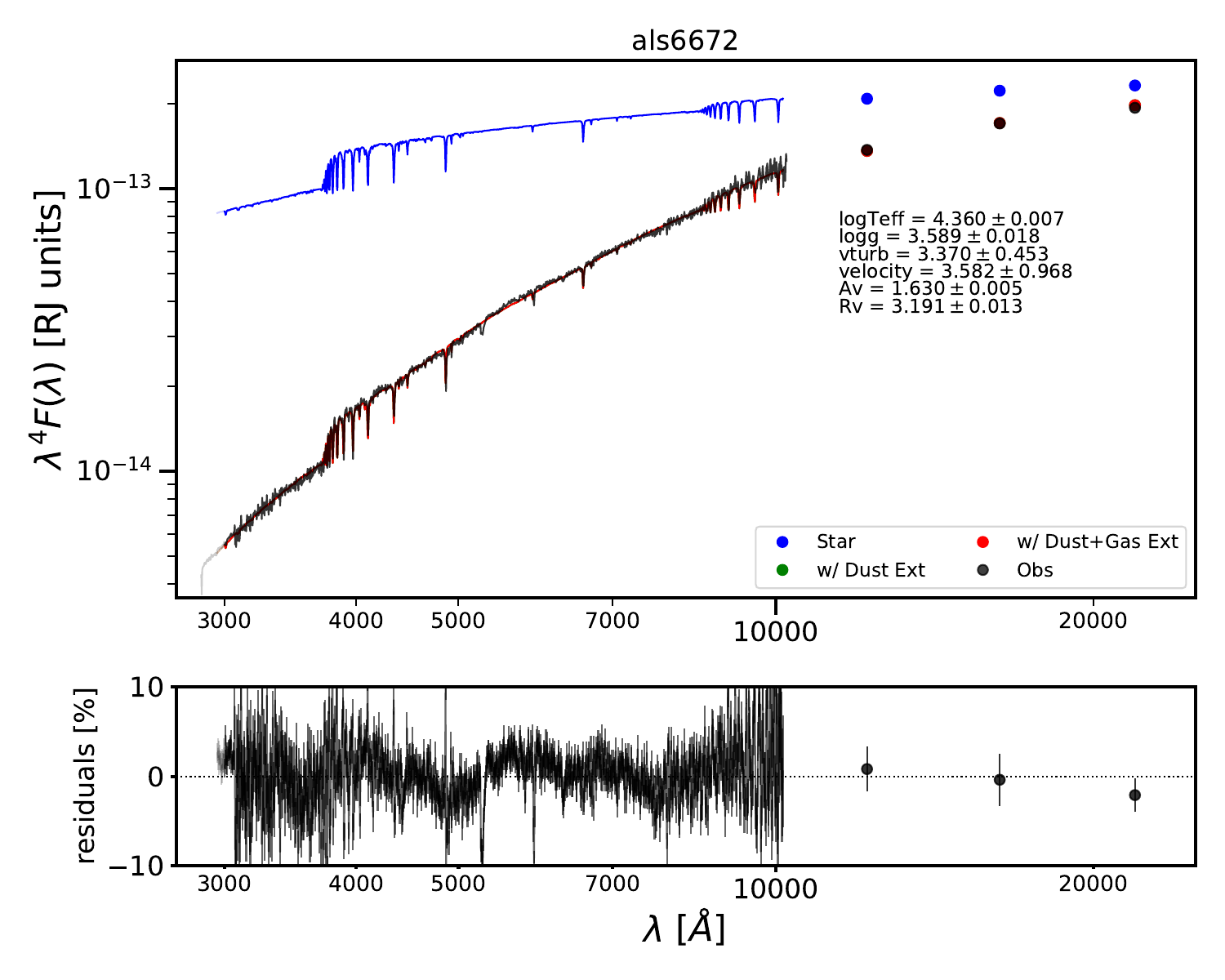}
      \caption{HST/STIS spectrum (black) of ALS6672, a B1V type star, fit with \texttt{measure\_extinction} package and TLUSTY using MCMC. The unreddened best-fit stellar atmosphere model is in blue, and the reddened model is in red. The three plot points on the right are the J, H, and K photometric bands.
      %\textcolor{red}{convert x-axis to non-science notation}
      }
    \label{fig:als6672_sed}
\end{figure*}
An example plot, for one of our fitted stellar spectra, ALS 6672, is provided in Figure~\ref{fig:als6672_sed}. This example illustrates the quality of the fit achieved by our method; the remaining targets are fitted to a comparable level. ALS 6672 was chosen as a representative case, lying between our best- and worst-fits.

\begin{figure}
  \centering
    \includegraphics[width=0.75\columnwidth]{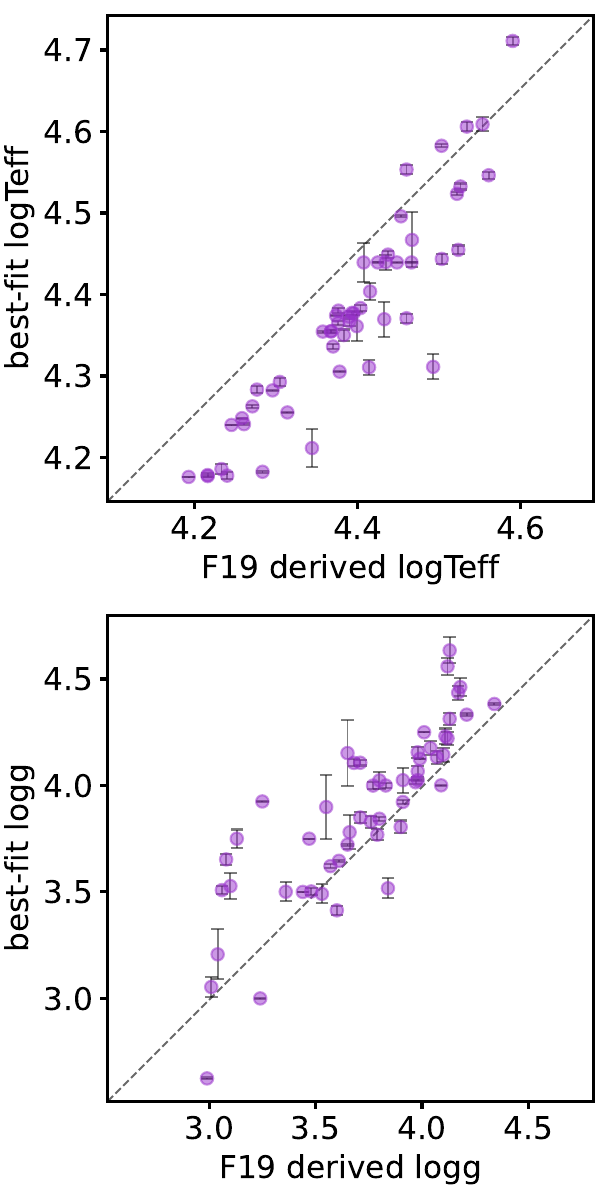}
      \caption{A comparison plot between the $\log{T_\mathrm{eff}}$, and $\log(g)$ derived in F19 using spline interpolation, and those derived in this paper. %These parameters were derived for the F19 sample of stars.
      }
    \label{fig:f19_compare}
\end{figure}
Figure\,\ref{fig:f19_compare} presents a scatter plot comparing the stellar parameters $\log{T_\mathrm{eff}}$ and $\log(g)$ from the above described method and the 2025 TLUSTY models, to those derived in F19. % using a spline interpolation method
The figure shows a systematic decrease in the best-fit stellar effective temperature, while finding a systematic increase in the best-fit surface gravity using the present method. {Due to the systematic nature of these trends, these differences between the present best-fit parameters and those appearing in F19 may be attributed to the updated stellar atmosphere models.}
%F19 used: y Lanz & Hubeny (2003, “OSTAR2002”) and Lanz & Hubeny (2007, “BSTAR2006”)

\subsection{Measuring the Extinction Curves}
\begin{figure*}
  \centering
  \includegraphics[width=1.0\textwidth]{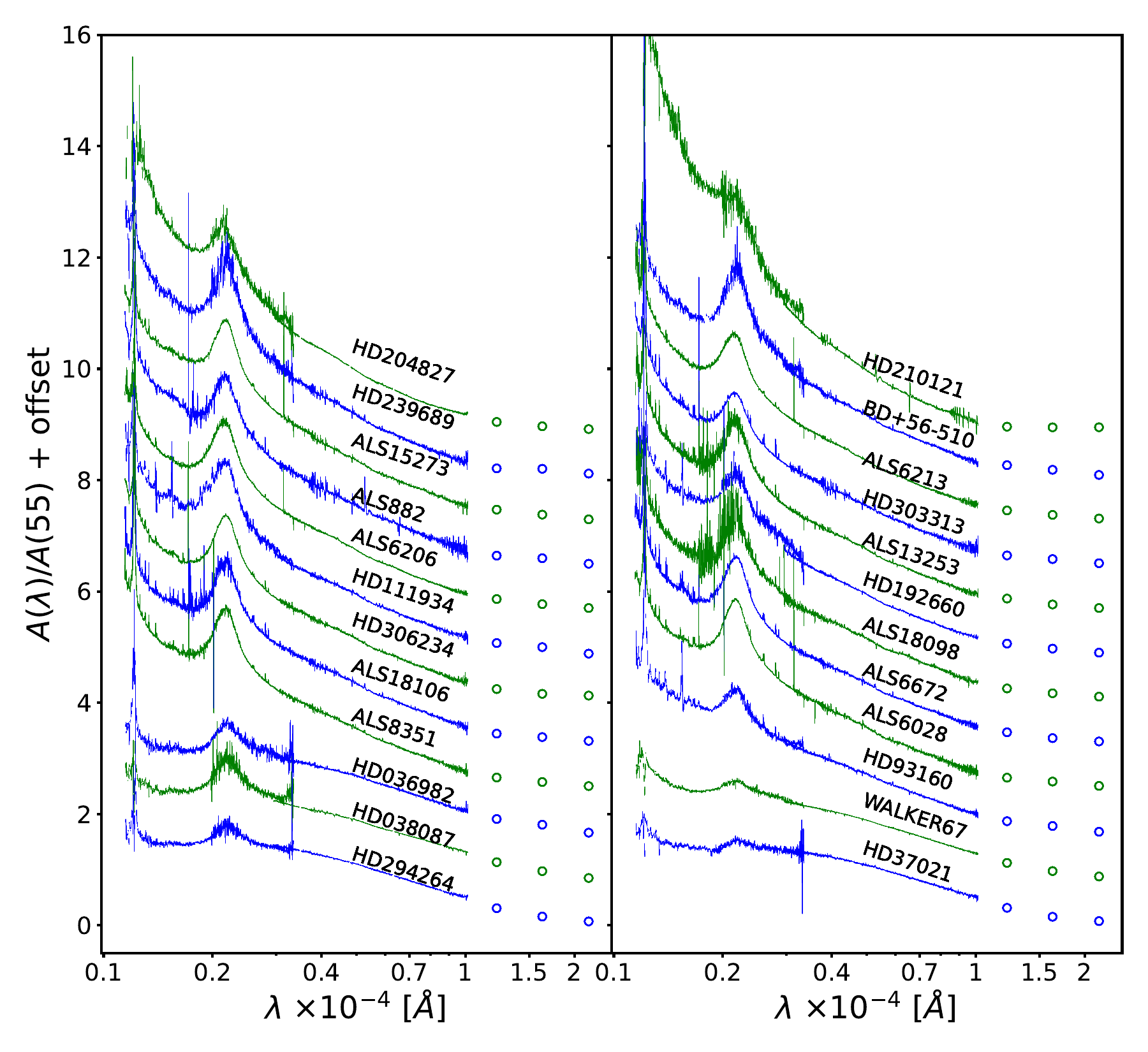}
      \caption{Measured UV and optical extinction curves of the \textit{Prince} sample. The curves are ordered by the overall slope of the extinction curve. The solid lines show the extinction curve spectral data, while the open circles give the JHK photometric points of the curve. The blue and green colors are simply used to distinguish between adjacent curves, and carry no particular meaning.}
    \label{fig:stacked_ext}
\end{figure*}
The extinction curves were then computed using the following procedure. To avoid the need for an accurate distance to the reddened and comparison stars, extinction is commonly measured relative to a reference wavelength. 
In order to maintain a consistent effective wavelength across different spectral types and dust abundances, as well as to mitigate any large uncertainties in the photometry, we compute extinction curves using the reference wavelength at $5500$\,\AA\ %and $4400$\,\AA\ 
in place of the traditional V band. Thus the dust extinction relative to $5500$\,\AA\ is given by,
\begin{equation}
    E(\lambda-55) = -2.5 \log{\frac{\mathcal{F}(\lambda)_{red}}{\mathcal{F}(\lambda)_{comp}}}+2.5 \log{\frac{\mathcal{F}(55)_{red}}{\mathcal{F}(55)_{comp}}}
    \label{eq:A55a}
\end{equation}
where, $\mathcal{F}(\lambda)_{red}$ is the flux of the reddened star, and $\mathcal{F}(\lambda)_{comp}$ is the flux of the comparison star from our best-fit stellar model. 
Then we convert $E(\lambda-55)$ to the absolute extinction $A(\lambda)$ normalized by the value at $5500$\,\AA\ using the relation,
\begin{equation}
    \frac{A(\lambda)}{A(55)} = \frac{E(\lambda-55)}{A(55)}+1.
    \label{eq:A55b}
\end{equation}
Figure~\ref{fig:stacked_ext} presents the measured UV and optical curves of the \textit{Prince} sample of stars.

Assuming that $A(\lambda)/A(55) \simeq A(\lambda)/A(V)$, we use reference values at J, H, K bands from \citet{2022ApJ...930...15D} in the relation,
\begin{equation}
    A(55) = \frac{E(\lambda-55)}{A(\lambda)/A(55) - 1}.
\end{equation}
This provides three independent measurements of $A(55)$. We take the uncertainty weighted average to obtain a final $A(55)$, effectively anchoring the extinction curve using the J, H, K bands.
The corresponding total-to-selective extinction at 5500\AA\ was then calculated with $R(55)=\frac{A(55)}{E(44-55)}$. The calculated $A(55)$ and $R(55)$ values for each line-of-sight are included in Table~\ref{tab:op_fit}. These results with $R(55)$ can be directly compared to works using $R(V)$, using the relation $R(55) = 0.99R(\rm V) - 0.049$ \citep{2019ApJ...886..108F}.

\subsection{Fitting the Extinction Curves}
\label{sec:fitting_ext}

\begin{deluxetable}{lccl}
\tablecaption{Optical extinction best-fit parameters.\label{tab:op_fit}}
\tablehead{
   \colhead{Column} & \colhead{Units} & \colhead{Label} & \colhead{Description}
}
\startdata
1 & \nodata & Starname & Stellar target \\
2 & {mag} & A(55) & Total extinction at {5500\,\AA} \\
3 & {mag} & $\sigma_{A(55)}$ & Total extinction at {5500\,\AA} uncertainty \\
4 & \nodata & R(55) & Total-to-selective extinction at {5500\,\AA} \\
5 & \nodata & $\sigma_{R(55)}$ & {Total-to-selective extinction uncertainty} at {5500\,\AA} \\
6 & {$A(\lambda)/A(55)$} & E$_0^{A(55)}$ & Curve intercept \\ 
7 & {$A(\lambda)/A(55)$} & $\sigma_{E_0^{A(55)}}$ & Intercept uncertainty \\
8 & {$A(\lambda)/A(55)\times{\rm \mu m}^{-1}$} & E$_1^{A(55)}$ & Curve slope \\ 
9 & {$A(\lambda)/A(55)\times{\rm \mu m}^{-1}$} & $\sigma_{E_1^{A(55)}}$ & Slope uncertainty \\
10 & {$A(\lambda)/A(55)\times{\rm \mu m}^{-2}$} & E$_2^{A(55)}$ & Quadratic coefficient \\ 
11 & {$A(\lambda)/A(55)\times{\rm \mu m}^{-2}$} & $\sigma_{E_2^{A(55)}}$ & Quadratic uncertainty \\
12 & {$A(\lambda)/A(55)\times{\rm \mu m}^{-3}$} & E$_3^{A(55)}$ & Cubic coefficient \\ 
13 & {$A(\lambda)/A(55)\times{\rm \mu m}^{-3}$} & $\sigma_{E_3^{A(55)}}$ & Cubic uncertainty \\
14 & {$A(\lambda)/A(55)\times{\rm \mu m}^{-4}$} & E$_4^{A(55)}$ & Leading coefficient \\ 
15 & {$A(\lambda)/A(55)\times{\rm \mu m}^{-4}$} & $\sigma_{E_4^{A(55)}}$ & Leading uncertainty \\
 & $A(\lambda)/A(55)$ & $\frac{A(\rm ISSnn)}{A(55)}$ & ISSnn feature amplitude \\
 & $A(\lambda)/A(55)$ & $\sigma_{A(\rm ISSnn)/A(55)}$ & ISSnn amplitude uncertainty \\
16-102 & \micron$^{-1}$ & x$_0^{\rm ISSnn}$ & ISSnn feature central wavelength \\
 & \micron$^{-1}$ & $\sigma_{{\rm x}_0^{\rm ISSnn}}$ & ISSnn central wavelength uncertainty \\
 & \micron$^{-1}$ & $\gamma_{\rm ISSnn}$ & ISSnn feature width \\
 & \micron$^{-1}$ & $\sigma_{\gamma_{\rm ISSnn}}$ & ISSnn width uncertainty \\
\enddata
\tablecomments{Table~\ref{tab:op_fit} is published in its entirety in the electronic edition of the {\it Astrophysical Journal}. The above description lists the columns in the table. %A portion is shown here for guidance regarding its form and content. 
The amplitude, central wavelength and width are included for all 17 features presented in Table~\ref{tab:iss}. Columns 16--102 are repeating sets of the amplitude, central wavelength and width, along with their respective uncertainties, for each of the 17 features presented in Table~\ref{tab:iss}.}
\end{deluxetable}

M20 showed that the optical extinction curve is well modelled by combining a low-order polynomial for the continuum with a Drude profile $D(x,x_i,\gamma_i)$ to capture each ISS feature,
\begin{equation}
\begin{split}
    \frac{A(\lambda)}{A(55)} =  \sum_{j=0}^{N_p}&E_j^{A(55)}x^{j} + \sum_{i=1}\frac{A(\rm ISS_i)}{A(55)}D(x,x_i,\gamma_i),\\ 
    &x \equiv 1/\lambda \\
    3000 {\rm\ \AA} &\leq \lambda \leq 10000 {\rm\ \AA}
\end{split}
\label{eq:4d_plus_drudes}
\end{equation}
in which the ISS features are represented by the Drude term,
\begin{equation}
\begin{split}
    D(x,x_o,\gamma)=\frac{x^2\gamma^2}{(x^2-x_o^2)^2+(x\gamma)^2},
\end{split}
\end{equation}
where $E_j^{A(55)}$ is the coefficient of each polynomial degree, $\frac{A(\rm ISS_i)}{A(55)}$ is the normalized strength of the ISS feature, and $\lambda_i,\gamma_i$ are the central wavelength and full-width-half-max (FWHM) of the ISS feature, respectively.
They also find that a fourth-order ($N_p=4$) polynomial effectively represents the overall shape of the optical curves, with higher-order polynomials providing no improvements in the fit quality. 
We use the superscript in the fit coefficients (both UV and optical) to distinguish the values derived from curves normalized by $A(55)$ from the standard normalization by $E(B-V)$ as in the FM90 parameters.

For the UV portion of the extinction curve, the FM90 parameterization is used, with the functional form \citep{2024ApJ...970...51G},
\begin{equation}
\begin{split}
    \frac{A(\lambda)}{A(55)} = C_1^{A(55)} &+ C_2^{A(55)}x + B_3^{A(55)}D(x,x_o,\gamma) \\&+ C_4^{A(55)}F(x), \\
    %&x \equiv 1/\lambda \\
    1000 &{\rm\ \AA} \leq \lambda \leq 3000 {\rm\ \AA}
\end{split}
\end{equation}
where the $2175$\,\AA\ bump is represented by the same form of the Drude term used for the ISS features above;
%\begin{equation}
%\begin{split}
%    D(x,x_o,\gamma)=\frac{x^2\gamma^2}{(x^2-%x_o^2)^2+(x\gamma)^2}
%\end{split}
%\end{equation}
where, $C_1^{A(55)}$, $C_2^{A(55)}$ and $C_4^{A(55)}$ are the intercept, slope and amplitude of the far-UV rise of the curve; the Drude $D(x,x_o,\gamma)$ models the $2175$\,\AA\ UV bump, where $x_o$ is the central wavelength of the bump and $B_3^{A(55)}$ is bump strength; lastly, the far-UV curvature is given by $F(x)$, which is a cubic function. %Note that, $F(x)$ here differs from $F_i^{A(55)}$ which are the ISS amplitudes. As mentioned above, the ${A(55)}$ superscript indicates that these parameter differ from the usual FM90 parameters which correspond to a curve normalized by $E(B-V)$.

Similar to the fitting described in the previous section, the fitting is done in two steps: 1. DLS fitting %with \texttt{LevMarLSQFitter}
2. MCMC fitting. %with \texttt{emcee}. 
For the latter step we used 5000 steps, and a burn fraction of 20$\%$, and again the number of walkers is twice the number of model dimensions. 
Same as before, the $5000$ steps did not satisfy the 50 times the autocorrelation time criterion, so we extended the chain length to $4\times10^4$ steps. However, this only changed the fitted parameter uncertainties by $\sim1$\% on average. So we determined a chain length of $5000$ steps provides adequate convergence.

%All scripts used to perform the extinction curve calculations and fitting, as well as to plot the results, are available in the \href{https://github.com/Chamani8/hst_iss_ext}{Chamani8/hst\_iss\_ext} \texttt{GitHub} repository. 
\begin{table*}
\centering
\caption{UV extinction fit parameters for the \textit{Prince} sample of stars.\label{tab:uv_fit_gun25}}
\begin{tabular}{lcccccc}
\hline\hline
Starname & C$_1$ & C$_2$ & B$_3$ & C$_4$ & x$_o$ (\micron$^{-1}$) & $\gamma$ (\micron$^{-1}$) \\
\hline
ALS882 & $0.714 \pm 0.009$ & $0.340 \pm 0.002$ & $1.163 \pm 0.007$ & $0.147 \pm 0.003$ & $4.569 \pm 0.003$ & $0.917 \pm 0.011$ \\
ALS1795 & $0.774 \pm 0.006$ & $0.314 \pm 0.001$ & $0.938 \pm 0.005$ & $0.225 \pm 0.002$ & $4.579 \pm 0.002$ & $0.896 \pm 0.010$ \\
ALS2285 & $1.018 \pm 0.005$ & $0.229 \pm 0.001$ & $1.275 \pm 0.004$ & $0.189 \pm 0.002$ & $4.555 \pm 0.001$ & $0.932 \pm 0.006$ \\
ALS6028 & $1.034 \pm 0.006$ & $0.238 \pm 0.001$ & $1.315 \pm 0.005$ & $0.171 \pm 0.002$ & $4.588 \pm 0.002$ & $0.953 \pm 0.007$ \\
ALS6206 & $0.966 \pm 0.005$ & $0.249 \pm 0.001$ & $1.282 \pm 0.007$ & $0.170 \pm 0.003$ & $4.618 \pm 0.002$ & $1.075 \pm 0.010$ \\
ALS6213 & $0.817 \pm 0.005$ & $0.315 \pm 0.001$ & $1.128 \pm 0.005$ & $0.202 \pm 0.003$ & $4.620 \pm 0.002$ & $1.005 \pm 0.008$ \\
ALS15273 & $0.856 \pm 0.006$ & $0.323 \pm 0.001$ & $1.310 \pm 0.005$ & $0.104 \pm 0.002$ & $4.580 \pm 0.002$ & $0.938 \pm 0.007$ \\
ALS6672 & $0.847 \pm 0.006$ & $0.276 \pm 0.001$ & $1.288 \pm 0.005$ & $0.199 \pm 0.004$ & $4.569 \pm 0.002$ & $0.937 \pm 0.007$ \\
ALS8351 & $1.044 \pm 0.005$ & $0.212 \pm 0.001$ & $1.229 \pm 0.005$ & $0.192 \pm 0.003$ & $4.562 \pm 0.002$ & $0.931 \pm 0.008$ \\
ALS13253 & $0.940 \pm 0.008$ & $0.277 \pm 0.002$ & $1.096 \pm 0.014$ & $0.224 \pm 0.006$ & $4.500 \pm 0.001$ & $0.823 \pm 0.015$ \\
ALS18106 & $1.054 \pm 0.009$ & $0.220 \pm 0.002$ & $1.167 \pm 0.011$ & $0.145 \pm 0.004$ & $4.502 \pm 0.002$ & $0.831 \pm 0.013$ \\
ALS18098 & $1.154 \pm 0.012$ & $0.224 \pm 0.003$ & $0.798 \pm 0.016$ & $0.262 \pm 0.008$ & $4.500 \pm 0.001$ & $0.803 \pm 0.003$ \\
BD+56-510 & $0.991 \pm 0.016$ & $0.297 \pm 0.003$ & $1.399 \pm 0.022$ & $0.202 \pm 0.005$ & $4.536 \pm 0.004$ & $0.922 \pm 0.020$ \\
HD036982 & $1.344 \pm 0.015$ & $0.022 \pm 0.003$ & $0.584 \pm 0.012$ & $0.116 \pm 0.004$ & $4.513 \pm 0.008$ & $0.834 \pm 0.023$ \\
HD037021 & $1.333 \pm 0.016$ & $-0.002 \pm 0.005$ & $0.185 \pm 0.010$ & $0.108 \pm 0.015$ & $4.512 \pm 0.011$ & $0.964 \pm 0.097$ \\
HD038087 & $1.340 \pm 0.013$ & $0.029 \pm 0.002$ & $0.706 \pm 0.014$ & $0.078 \pm 0.003$ & $4.520 \pm 0.007$ & $0.950 \pm 0.027$ \\
HD093160 & $0.426 \pm 0.007$ & $0.366 \pm 0.002$ & $0.495 \pm 0.006$ & $0.008 \pm 0.004$ & $4.895 \pm 0.005$ & $1.434 \pm 0.035$ \\
HD111934 & $0.782 \pm 0.008$ & $0.285 \pm 0.002$ & $1.396 \pm 0.011$ & $0.147 \pm 0.005$ & $4.630 \pm 0.003$ & $1.149 \pm 0.016$ \\
HD192660 & $0.910 \pm 0.018$ & $0.296 \pm 0.003$ & $0.922 \pm 0.017$ & $0.103 \pm 0.008$ & $4.557 \pm 0.006$ & $1.201 \pm 0.042$ \\
HD204827 & $0.515 \pm 0.013$ & $0.472 \pm 0.002$ & $1.096 \pm 0.029$ & $0.006 \pm 0.007$ & $4.558 \pm 0.006$ & $0.897 \pm 0.027$ \\
HD210121 & $-0.446 \pm 0.016$ & $0.858 \pm 0.004$ & $0.796 \pm 0.028$ & $0.263 \pm 0.010$ & $4.602 \pm 0.013$ & $0.901 \pm 0.043$ \\
HD239689 & $0.771 \pm 0.019$ & $0.355 \pm 0.003$ & $1.505 \pm 0.040$ & $0.195 \pm 0.007$ & $4.544 \pm 0.006$ & $0.954 \pm 0.031$ \\
HD294264 & $1.287 \pm 0.008$ & $0.018 \pm 0.001$ & $0.379 \pm 0.011$ & $0.067 \pm 0.002$ & $4.510 \pm 0.007$ & $1.067 \pm 0.043$ \\
WALKER67 & $1.147 \pm 0.004$ & $0.068 \pm 0.001$ & $0.315 \pm 0.004$ & $0.115 \pm 0.002$ & $4.503 \pm 0.003$ & $1.242 \pm 0.029$ \\
\hline
\end{tabular}
\end{table*}

\begin{figure*}
  \centering
  \includegraphics[width=0.7\textwidth]{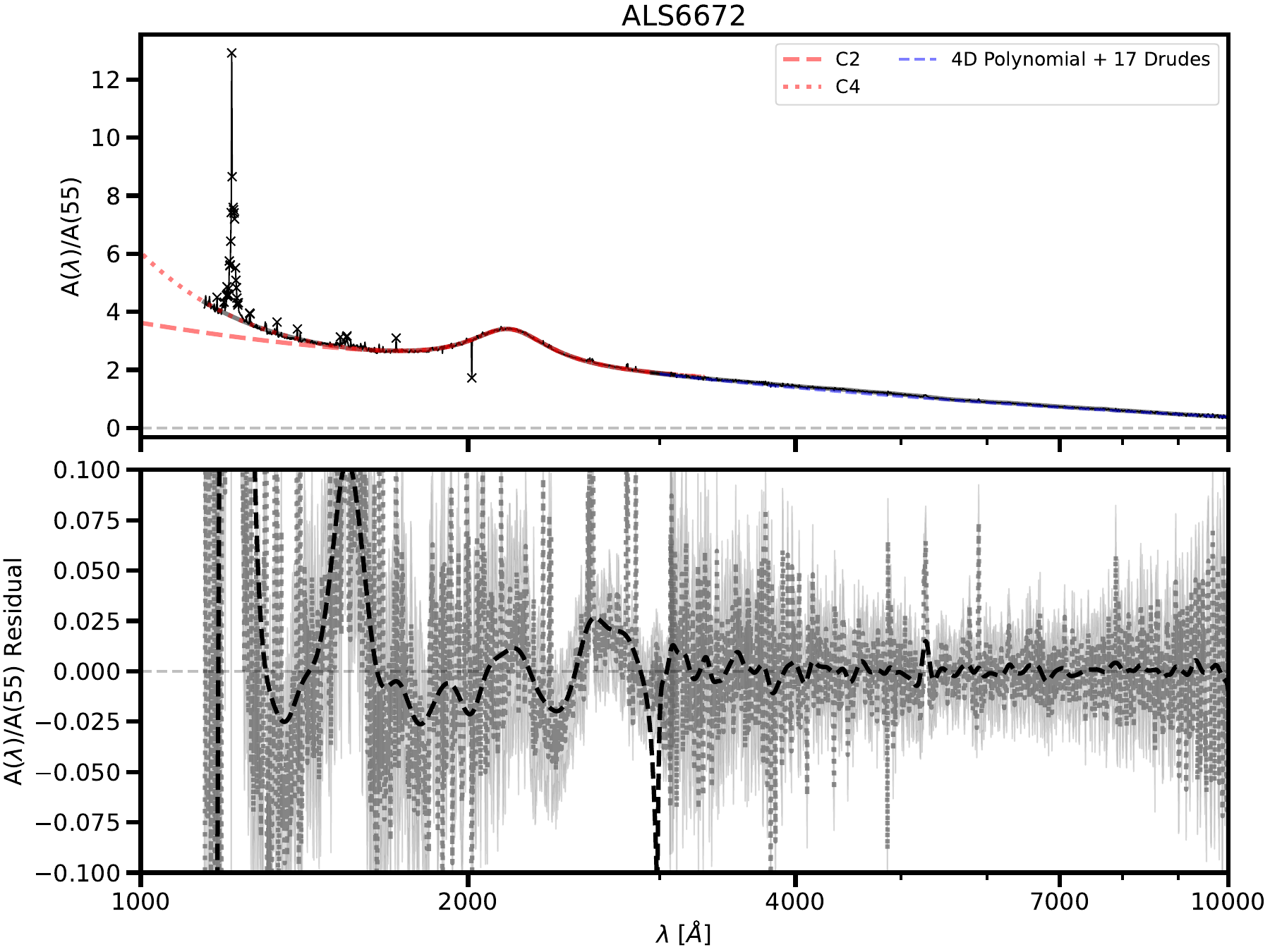}
      \caption{Measured (black) and fitted (red and blue) extinction curve for ALS6672, using the FM90 parameterization in the UV, and a 4D polynomial with Drude profiles in the optical. The ``x" points represent outlying points excluded from the fit. The residuals are shown in the bottom panel, with the dashed line representing the residuals convolved with a Gaussian filter of width 150\,\AA.}
    \label{fig:als6672_ext}
\end{figure*}

Figure~\ref{fig:als6672_ext} shows an example plot of the fitted extinction curve using the above described method, illustrating the quality of the fitting method. The best-fit parameters of the UV and optical extinction curves are summarised in {Table~\ref{tab:uv_fit_gun25} and \ref{tab:op_fit}}, respectively. %Only a subset of the Table~\ref{tab:op_fit} is shown here; the full table is available in the online supplementary material. 
These parameters are analyzed in the following section to identify and characterise all the observed ISS features.

\section{Analysis of ISS features}
\label{sec:analysis}

\begin{figure*}
  \centering
  \includegraphics[width=0.6\textwidth]{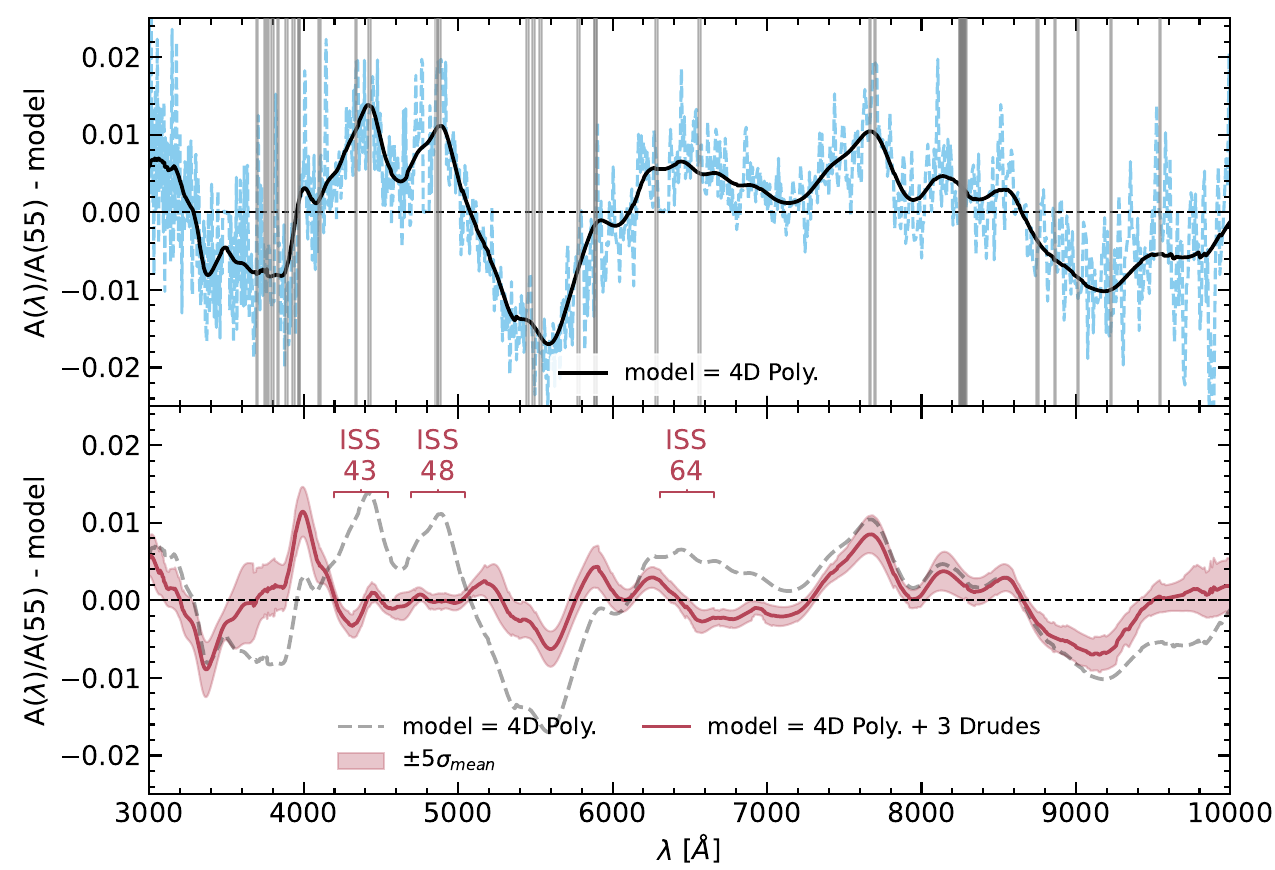}
      \caption{Residuals of fitting the optical extinction curves. 
      \textit{Top:} The blue dashed plots show the residuals of all 74 targets at the full resolution. The black solid plot is the mean of the residuals convolved with a Gaussian filter, resolving only features wider than $150$\,\AA. The vertical grey lines represent the interstellar lines and DIBs that were smoothed over. \textit{Bottom:} The red solid plot represents the re-binned mean residual of extinction curves fitted with Drude profiles for the three original ISS features. The fitted ISS features are indicated by the red labels, ``ISSnn''. The shaded area around the red solid plot, represents the 5 times the standard deviation of the mean residual. The grey dashed line is the same re-binned mean residual as shown in the top panel; included here for comparison.
      }
    \label{fig:stacked_orig_iss}
\end{figure*}

Unlike narrow features such as DIBs, analysing features with widths larger than $150$\,\AA\ presents a significant challenge. 
The true start and end of the faint ISS feature can be conflated with nearby overlapping features and noise in the data. 
Following the analysis method used in M20 to overcome these difficulties, we adopt the approach of analysing the magnitudes of the extinction curve residuals. The residuals (Eq.~\ref{eq:4d_plus_drudes} subtracted from the measured extinction curve) are obtained by subtracting the measured extinction curves obtained in Section\,\ref{sec:measuring_ext} by the best-fit models obtained in Section\,\ref{sec:fitting_ext}. {Additionally, as there has been no evidence of DIBs or ISS features in the UV or far-UV \citep{2003ApJ...592..947C, 2009ApJ...705.1320G}{, and as we likewise find no such features in our own data}, we limit our analysis to the Optical (see further discussion in Section~\ref{sec:summary}).}
%\todo{Q3. explain that we limit confirming + finding new features to the optical, as all the ISS features that have been found thus far have been in the optical. For now we look for ISS features = broad features in the optical.}

%\begin{equation}
%    r(\lambda - 55) = \frac{A(\lambda)}{A(55)} -  %(\sum_{j=0}^{N_p}E_jx^j + \sum_{i=1}F_i^{A(55)}D(x,x_i,\gamma_i)).
%\end{equation}
%\todo{this equation maybe not needed}
We begin with a featureless model and iteratively add ISS features based on the residuals.
The top panel of Figure\,\ref{fig:stacked_orig_iss} presents the residuals obtained by fitting the optical extinction curves with only a fourth degree (4D) polynomial. First we fit the extinction curve of each target as described in Section\,\ref{sec:fitting_ext}, of which the residuals at full resolution of the data are shown by the blue dashed lines. To ensure that our analysis only focuses %centres 
on the ISS features, we interpolate over the most prominent interstellar and hydrogen lines, and DIBs in this region of the spectrum (gray vertical lines in Figure\,\ref{fig:stacked_orig_iss}, top panel). %As can be seen from the multitude of ``wiggles'' in these dashed plots, it is difficult to distinguish the ISS features from the noise at this resolution. 
To highlight the broad features and improve their S/N, the black line shows the residual convolved with a Gaussian, retaining only features wider than $150$\,\AA, and suppressing narrower ones. %the black line is the mean of the residuals $r(\lambda-55)$ convolved with a Gaussian filter of width to retain only features wider than $150$\,\AA, suppressing narrower features. 
This width limit of $150$\,\AA, was chosen to confidently exclude any known DIBs, of which the widest one, at $\sim4428$\,\AA\, has an FWHM of $\sim10-30$\,\AA\ \citep{1963ApJ...137..200H, 2020AJ....159..113G}.  
%The full list is provided in Table\,\ref{tab:lines}. 
Lastly, we mask any outlying points from each individual extinction curve\footnote{{Masking was done using the \texttt{sigma\_clip} routine from \texttt{astropy}.}}, before computing the mean of the remaining residuals. %The mean residual obtained by only fitting the 4D polynomial is shown in the top panel of Figure\,\ref{fig:stacked_orig_iss}. In blue is the mean residual at the full HST/STIS resolution.

The bottom panel of Figure\,\ref{fig:stacked_orig_iss} presents the mean residual with the inclusion of the three original ISS features discovered by M20, namely ISS43, ISS48 and ISS64, in the fitting. We have chosen to name the ISS features using the first two significant figures of their central wavelength. 
For reference, the mean residual from the 4D polynomial-only fit is shown in the background using a dashed line with reduced opacity. We shade in five times the standard deviation of the mean of the residuals, showing that the remaining structure in the residuals %features observed
is stronger than the error.
Best-fit central wavelengths, widths, and amplitudes of all the ISS features (including the candidate ISS features that will be presented in Section~\ref{sec:new_iss}) are given in Table~\ref{tab:op_fit}. 
We see that a great deal of structure remains in these residuals. So in the following sections we analyse the newly published features ISS77, ISS54, ISS84, and propose an additional {9} new candidate ISS features.

\subsection{ISS features at 8500\,\AA\ and 7700\,\AA}

\begin{figure*}
  \centering
  \includegraphics[width=0.7\textwidth]{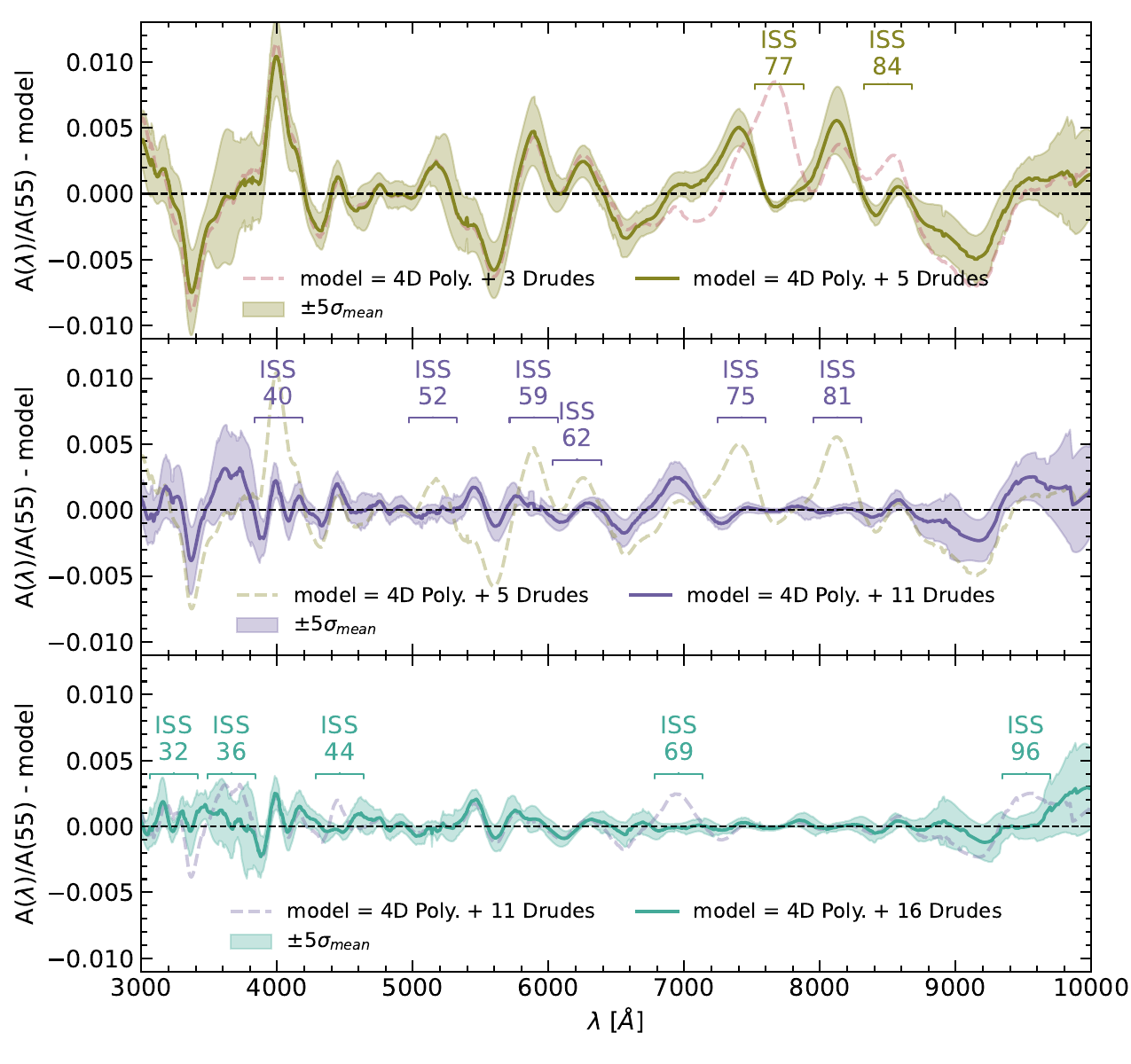}
      \caption{Average of the residuals of the extinction curves, fitted with the 2 literature-found (top panel) and {11} new (bottom two panels) candidate ISS features. The solid lines represent the averaged residuals resulting from fitting the indicated ISS features. For comparison, the less opaque dashed line represents the residual without the labelled features in the fitting--these also correspond to the residual in the panel above.}
    \label{fig:stacked_new_iss}
\end{figure*}

Beyond the ones identified in M20, M21 reveals a candidate ISS feature at 7700\,\AA\ and G25 reveals one at 8500\,\AA, detected using low-resolution Gaia spectra. To verify these two ISS features, we present the top panel in
Figure\,\ref{fig:stacked_new_iss}. 
The green plot (top panel) shows the mean residual after fitting the original three ISS features, and the features at 7700\,\AA\ and at 8500\,\AA. As before, a reference dashed plot with reduced opacity is included in each panel of the figure, which excludes the relevant ISS features in the extinction curve fitting. The central wavelengths and widths are fitted and not fixed at each step of this analysis. 
As evident by the difference between the two %red plot dashed vs. green solid 
lines in the top panel of Figure\,\ref{fig:stacked_new_iss}, there is an %significant 
improvement to the mean residual by including these two ISS features in the fitting. This demonstrates that the model incorporating the 7700\,\AA\ and 8500\,\AA\ better captures the shape of the observed extinction curve. 
Although these features are faint, the consistent impact on the fits by averaging across {74} targets, suggests they are not a result of random noise or artifacts, but rather real components of the curve. This provides strong empirical support for the presence of features ISS77 and ISS84 in these MW extinction curves. 

% that were previously undetectable using lower resolution data, such as that from Gaia. 

\subsection{A special case: the feature at 5400\,\AA}

\begin{figure}
    \centering
    \includegraphics[width=1.0\linewidth]{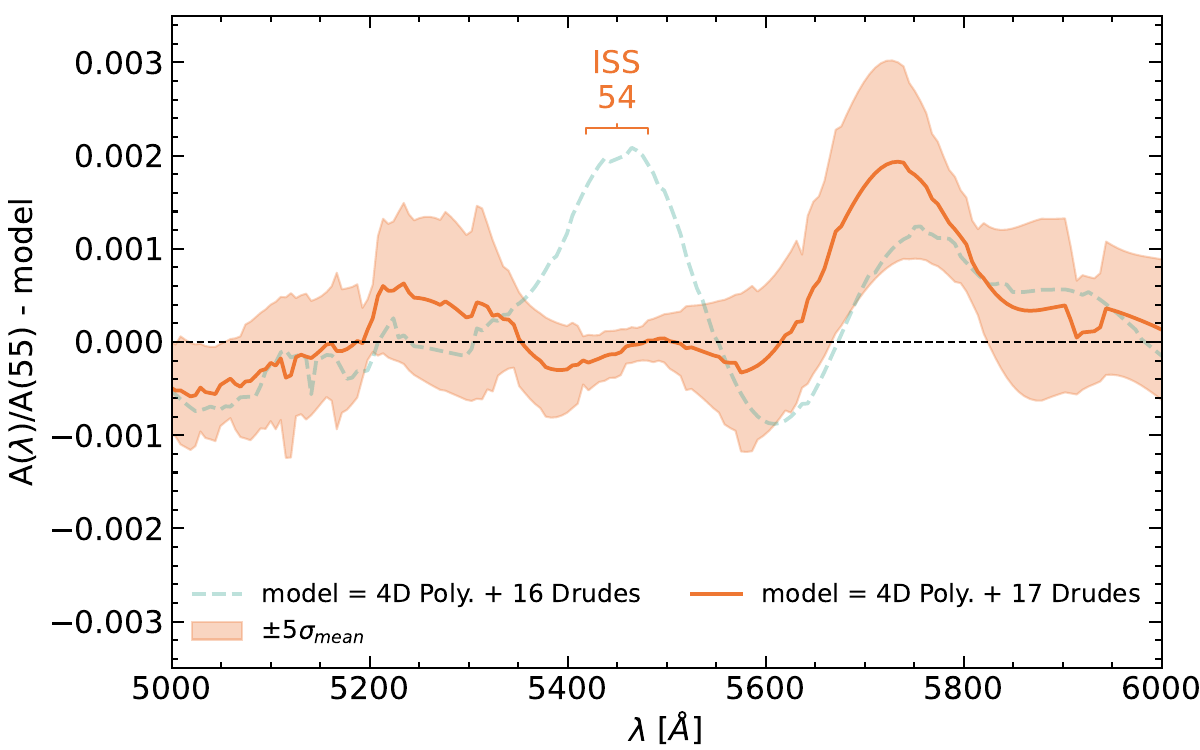}
    \caption{The average residual of fitting the ISS feature at 5400\,\AA\ feature in orange. For comparison we include the same residual appearing at the bottom panel of Figure~\ref{fig:stacked_new_iss}, which includes fits to the literature and candidate ISS features presented in Section~\ref{sec:new_iss}, except for ISS54, in teal.}
    \label{fig:iss5}
\end{figure}

In this section, we discuss the last literature published ISS feature found at 5400\,\AA\ in Z24. Figure~\ref{fig:iss5} shows the average residual when including a Drude at 5400\,\AA\ in our optical extinction curve model. This figure shows that the average residual is improved by including the ISS54 feature. 
However, Table~\ref{tab:iss} shows that we only find a significance of {3.2 for ISS54}. Consequently, this result does not meet the 5$\sigma$ criterion for a statistically significant detection, and we therefore cannot confirm the detection of 5400\,\AA\ feature for this particular sample of extinction curves.

In the following subsection, we present the remaining structure in the residuals of the extinction curves as new candidate ISS features.

\subsection{New Candidate ISS features}
\label{sec:new_iss}

Even with ISS features published in the literature removed from the mean extinction curve residual we find persisting structure larger than 5$\sigma$ for this sample of MW curves, as seen by the green plot (top panel) of Figure~\ref{fig:stacked_new_iss}. Adding in Drude profiles at {4010, 5150, 5890, 6210, 7420, and 8130}\,\AA, we obtain the purple plot (middle panel) of Figure~\ref{fig:stacked_new_iss}. In this plot we find several more features that were previously difficult to see, so we add in Drudes at {3240, 3660, 4460, 6960, and 9520}\,\AA, finally giving us the flat residual in the bottom panel in teal. This results in a total of {11} candidate ISS features. A summary of the 6 literature and {11} candidate ISS are presented in Table\,\ref{tab:iss}.

To evaluate the S/N ratio of these candidate ISS features we compute their significance in the following manner. First we obtain the standard deviation of the mean (SDOM)\footnote{{We use the standard definition of SDOM as it appears in \citet{2003drea.book.....B}.}} of the mean extinction curve. 
Then with the SDOM as the uncertainty of this mean extinction curve, we conduct an MCMC fitting as described in Section~\ref{sec:fitting_ext} for all {17} features (the 6 literature ISS features and the {11} candidate features). Table~\ref{tab:iss} presents the resulting significances computed using the area under a Drude profile. As a check, the best-fit amplitude divided by the uncertainty yields the same significances as those computed using the area under the curve. Both significance calculations find that all but two of the ISS features, ISS54 and ISS62, are well above a 5$\sigma$ detection, providing them with strong statistical confidence.

\begin{table*}
\centering
\caption{ISS Feature Properties \label{tab:iss}}
\begin{tabular}{lccccccc}
\hline\hline
Feature & Family & $\lambda_o$ (${\rm \AA}$) & $\gamma$ (${\rm \AA}$) & Amplitude\footnote{This amplitude is in units of $A({\rm ISS})/A(55) \times 10^4$.} ($\times 10^4$) & Area\footnote{The area under the Drude profile, which was computed using the best-fit amplitude and width of the feature.}/$\sigma_{area}$ & Dust feature? & Origin\\
\hline
ISS32 & $\beta$  & $3249.9 \pm 2.1$  & $153.7 \pm 4.8$  & $70.6 \pm 7.0$   & 10.1 & No     & \textbf{new} \\
ISS36 & $\alpha$ & $3652.3 \pm 2.1$  & $594.9 \pm 1.9$  & $109.8 \pm 10.7$ & 10.3 & Yes    & \textbf{new} \\
ISS40 & $\alpha$ & $3993.6 \pm 1.5$  & $305.7 \pm 6.6$  & $181.3 \pm 5.7$  & 31.4 & Yes    & \textbf{new} \\
ISS43 & $\alpha$ & $4353.5 \pm 7.7$  & $504.1 \pm 8.1$  & $364.1 \pm 4.2$  & 84.6 & Yes    & \textbf{M20} \\
ISS44 & $\alpha$ & $4442.5 \pm 1.3$  & $65.2 \pm 1.7$   & $101.2 \pm 3.4$  & 29.2 & Yes    & \textbf{new} \\
ISS48 & $\alpha$ & $4847.3 \pm 2.8$  & $462.4 \pm 0.8$  & $383.3 \pm 5.4$  & 70.4 & Yes    & \textbf{M20} \\
ISS52 & $\alpha$ & $5176.0 \pm 9.5$  & $345.6 \pm 4.4$  & $111.6 \pm 5.4$  & 20.8 & Yes    & \textbf{new} \\
ISS54 & $\alpha$ & $5423.0 \pm 3.1$  & $107.5 \pm 1.2$  & $14.4 \pm 4.5$   & 3.2  & Unsure & \textbf{Z24} \\
ISS59 & $\beta$  & $5872.0 \pm 10.1$ & $209.8 \pm 1.7$  & $106.3 \pm 6.8$  & 15.7 & Yes    & \textbf{new} \\
ISS62 & $\beta$  & $6172.8 \pm 0.6$  & $215.1 \pm 3.8$  & $47.1 \pm 11.0$  & 4.3  & Unsure & \textbf{new} \\
ISS64 & $\beta$  & $6443.3 \pm 6.5$  & $1103.9 \pm 9.4$ & $321.5 \pm 6.3$  & 49.9 & Yes    & \textbf{M20} \\
ISS69 & $\beta$  & $6910.9 \pm 5.4$  & $279.0 \pm 17.0$ & $53.8 \pm 5.2$   & 10.0 & Yes    & \textbf{new} \\
ISS75 & $\beta$  & $7457.1 \pm 18.4$ & $541.2 \pm 2.6$  & $152.9 \pm 5.2$  & 29.1 & Yes    & \textbf{new} \\
ISS77 & $\beta$  & $7710.1 \pm 14.5$ & $191.8 \pm 4.2$  & $178.6 \pm 6.6$  & 25.4 & Yes    & \textbf{M21} \\
ISS81 & $\beta$  & $8090.6 \pm 19.4$ & $228.9 \pm 5.5$  & $142.7 \pm 10.1$ & 13.7 & Yes    & \textbf{new} \\
ISS84 & $\beta$  & $8431.7 \pm 4.7$  & $374.5 \pm 2.3$  & $121.7 \pm 3.7$  & 32.3 & Yes    & \textbf{Z24} \\
ISS96 & --       & $9569.4 \pm 6.8$  & $191.2 \pm 5.0$  & $53.9 \pm 5.2$   & 10.1 & No     & \textbf{new} \\
\hline
\end{tabular}
\end{table*}

\section{Results: ISS Correlations}
\begin{figure*}
  \centering
  \includegraphics[width=\textwidth]{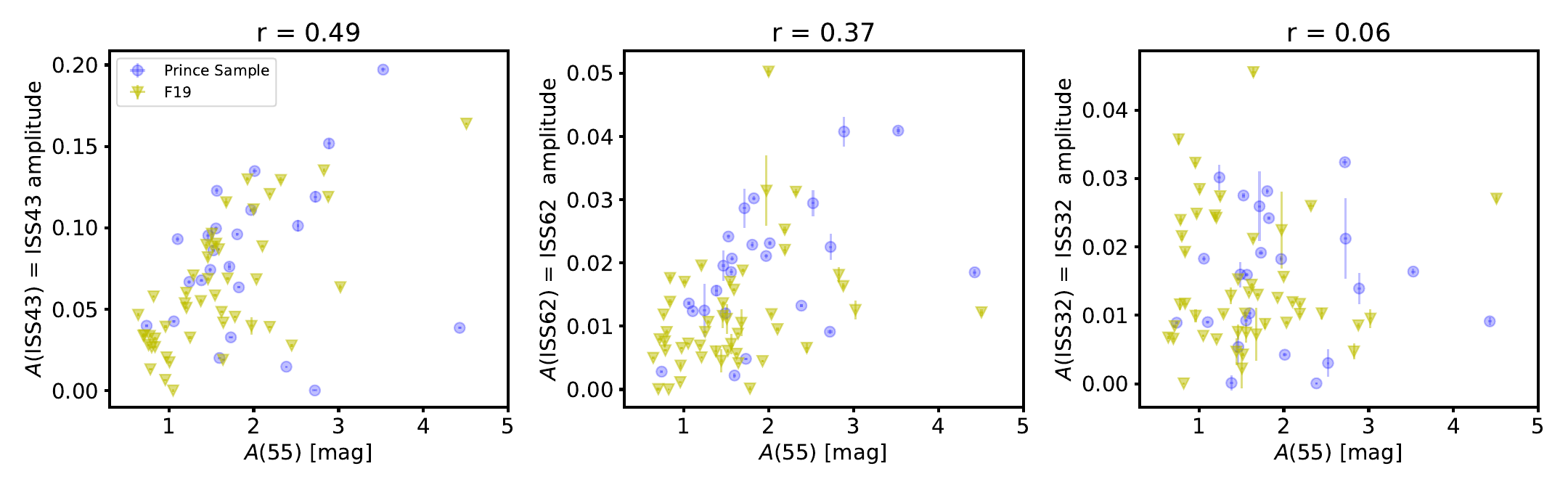}
  \includegraphics[width=\textwidth]{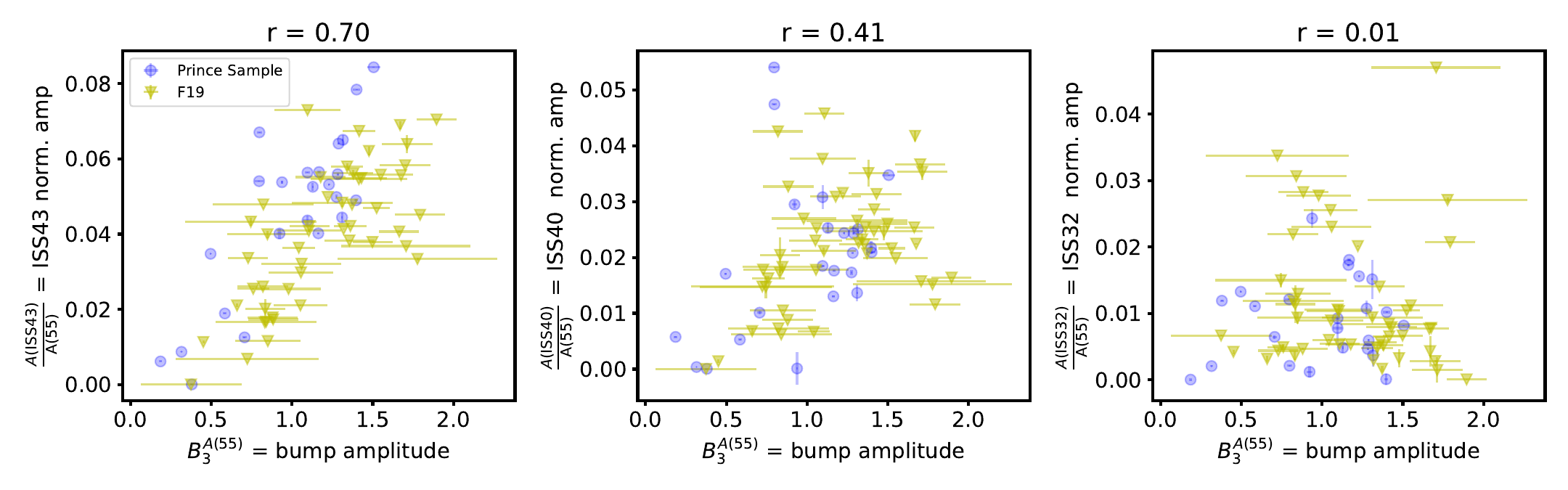}
  \includegraphics[width=\textwidth]{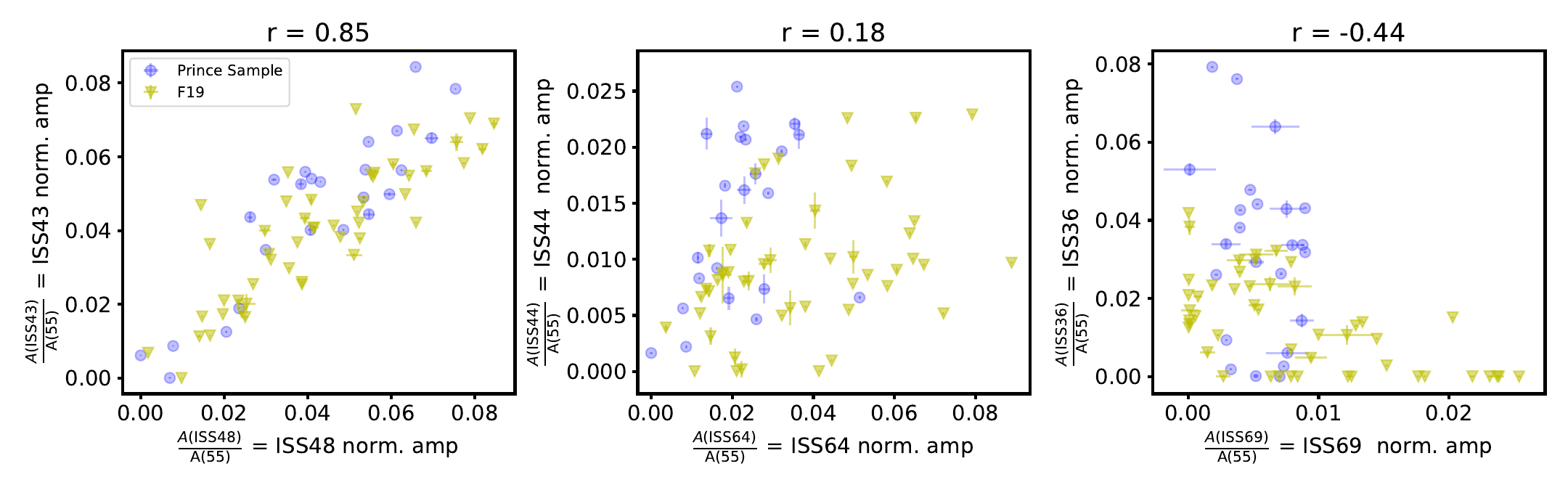}
      \caption{Example plots of amplitude of ISS features, plotted against total extinction, $A(55)$, (top row) and $2175$\,\AA\ UV bump amplitude, $B_3$ (middle row), and other ISS feature amplitudes (bottom row). Note that the y-axis in the top row is the un-normalized feature strength $A(\rm ISSnn)$, whereas the y-axis in the middle and bottom rows are the normalized feature strength $\frac{A(\rm ISSnn)}{A(55)}$, to remove the dependency on $A(55)$. Here, $r$ is the Pearson correlation coefficient of the corresponding plot. The left column of panels shows an example of an ISS feature that is strongly correlated with ${A(55)}$, $B_3^{A(55)}$, or another ISS feature, whereas the right panels show ones with the weakest or near-zero correlation. The right panel of the bottom row shows an example of an ISS feature that is anti-correlated with another feature.
      }
    \label{fig:iss_scatter}
\end{figure*}

%The primary aim of this investigation is to provide further insight into the nature of ISS features. 
ISS features are clearly observable in the optical portion of spectroscopic extinction curves. However, it is not apparent that the carriers of these features are interstellar dust. Given that these features are broader than any gas-phase absorption-line, their dust nature can be determined by examining if the strength of the features correlates with the dust column density i.e. $A(55)$. 
Furthermore, correlations among the feature strengths can reveal whether all the ISS features belong to the same %dust
carrier or multiple different carriers.
Lastly, correlations with other extinction parameters, such as $B_3$ (the $2175$\,\AA\ bump strength), $C_4$ (the far UV rise), or $R(55)$, %which carries information about the dust size distribution.
%can be used to probe further into the nature of the carrier(s).
provide additional constraints on the nature of the carrier(s) responsible for the ISS features.

Figure\,\ref{fig:iss_scatter} presents a few example plots of amplitude of the ISS features plotted against the total extinction, $A(55)$, in the top panels, the $2175$\,\AA\ bump strength, $B_3^{A(55)}$ in the middle row panels and other ISS features in the bottom row.  In the top row of panels in Figure\,\ref{fig:iss_scatter}, we compare the un-normalized ISS feature strengths $A(\rm ISSnn)$ with $A(55)$, allowing us to examine how the absolute feature strength correlates with the dust column density. In contrast, in the middle row of the same figure, we compare the feature strengths with the bump strength $B_3^{A(55)}$, both of which are normalized by $A(55)$ to remove the dependence on overall dust column density. This way we can identify trends between the ISS features and the bump, independent of mutual correlation with the dust column. 
The Pearson product-moment correlation coefficients ($r$) are presented above each plot\footnote{{Correlation coefficients were computed using the \texttt{corrcoef} function from \texttt{numpy}.}}. We provide these plots to illustrate what the correlations look like at different values of $r$, and leave the interpretation of the correlations to the remainder of the Section. The best-fit central wavelengths and widths of the features show little to no variation, and no correlations, so these parameters are not discussed in the remainder of this section. %The best-fit values of the central wavelengths and widths are provided in Table\,\ref{tab:op_fit}.

\begin{figure*}
    \centering
    \includegraphics[width=0.7\textwidth]{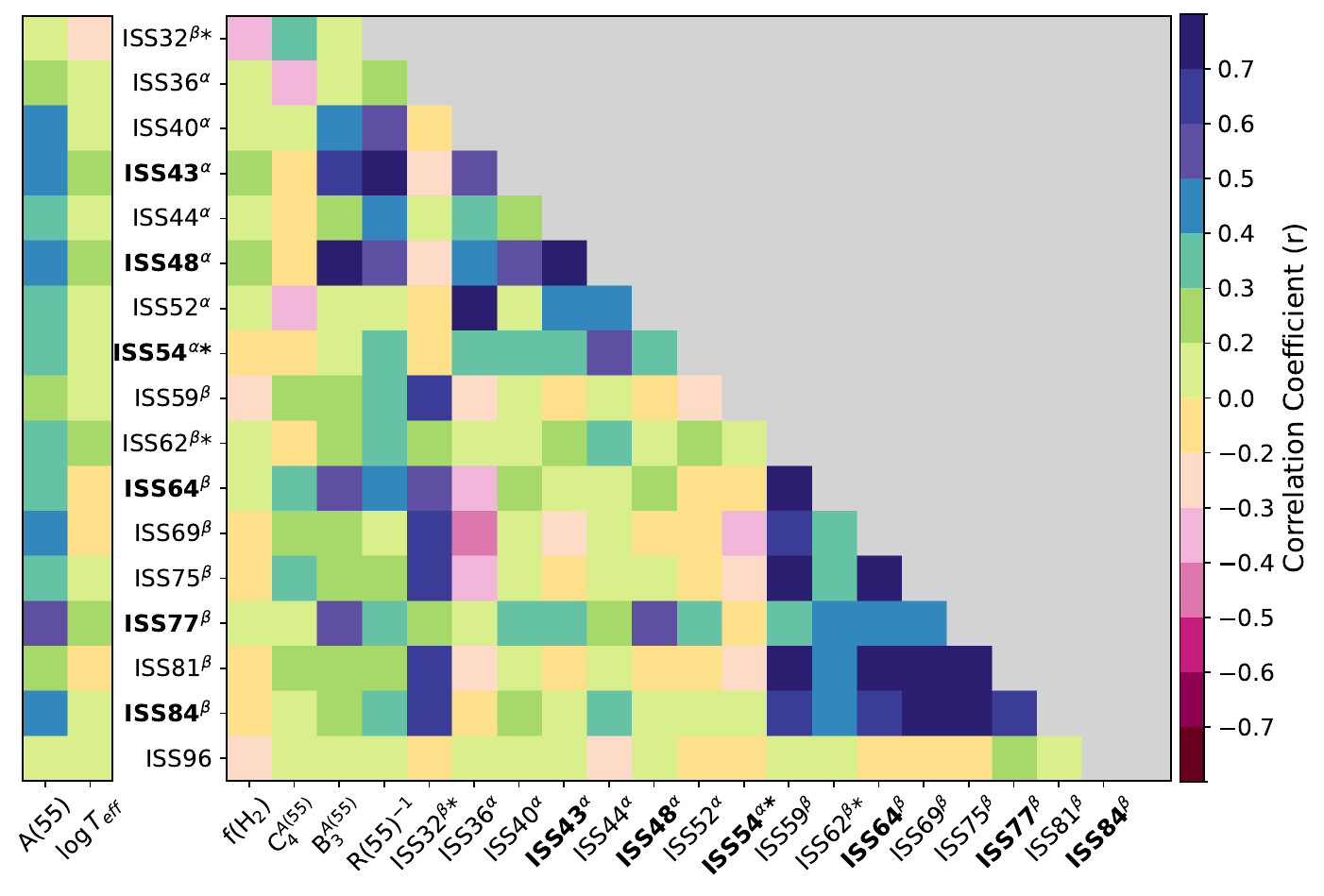}
    \caption{Colormap of the correlation coefficients between the feature strengths with A(55), 1/R(55), the bump strength ($B_3$), the far-UV rise ($C_4^{A(55)}$), log of the effective stellar temperature ($\log{T_{eff}}$), hydrogen column density $f({\rm H_2})$ and each other. The $\alpha$ and $\beta$ superscripts indicate the family of the ISS feature. The left panel presents the parameters correlated with $A(\rm ISSnn)$, while the right panel shows the parameters correlated with $\frac{A(\rm ISSnn)}{A(55)}$. The 6 literature ISS features are distinguished by bold font. Features
    labelled ``ISSnn*'' denote features that may not be dust features or have a low detection.}
    \label{fig:corr_map}
\end{figure*}
Given the 17 ISS features, we do not present all individual correlation scatter plots. Instead, Figure\,\ref{fig:corr_map} presents a colormap summarising the correlation coefficients between the ISS features amplitudes, the extinction parameters $A(55)$, $1/R(55)$ and $B_3^{A(55)}$, and the stellar parameter $\log T_\mathrm{eff}$. Here we show $A(55)$, $\log{T_{eff}}$ and $f({\rm H_2})$ in a separate column since these values are compared with $A(\rm ISSnn)$, instead of $\frac{A(\rm ISSnn)}{A(55)}$ as with $1/R(55)$, $C_4^{A(55)}$ and $B_3^{A(55)}$, eliminating the dependency on dust column. 
The molecular fraction, $f({\rm H_2})$ is computed using the linear relation presented in \citet{2023ApJ...944...33V}.
We interpret the correlation coefficients in the remainder of the paper as follows: $0.00\leq r \leq0.19$ no correlation, $0.2\leq r \leq0.29$ very weak correlation, $0.30\leq r \leq 0.49$ moderate correlation, $0.50\leq r \leq 0.69$ strong correlation, and $0.70\leq r \leq 1.00$ very strong correlation.

% Are they dust features?
The correlations in the left panel with $A(55)$ ($r>0.2$) reveal that the amplitudes of all ISS features except ISS32 and ISS96 are dust features. This is confirmed by the lack of correlations between the remainder of the feature strengths with $\log T_\mathrm{eff}$. As can be seen in the bottom row of the figure, ISS96 is not well correlated with any other features or parameters such as $\log{T_{eff}}$ and $A(55)$. Thus we do not consider it a candidate ISS feature, and instead it is likely a poorly fit stellar line or a systematic anomaly in the data. On the other hand, although ISS32 does not correlate with $A(55)$, it does have strong correlations with some other ISS features, and thus we retain it as a candidate ISS feature.
%{ISS32 and ISS96} are likely not well fit stellar lines in the stellar atmosphere model or systematic anomalies in the data. 
This leaves us with a total of 9 new candidate ISS features, of the 10 statistically significant new features.

\section{Discussion}
\label{sec:discussion}
%correlations with other parameters
% Correlation scale:
%0.00–0.29	very weak/None
%0.30–0.49	Moderate    (teal-blue)
%0.50–0.69	Strong      (purple)
%0.70–1.00	Very strong (darkest purple)
Only ISS43, ISS48, ISS64 and ISS77 have a strong positive correlation ($r>0.5$) with the strength of the $2175$\,\AA\ bump ($B_3$), indicating that {they arise from the same environments and that their carrier populations are correlated. The strong evidence that carbonaceous grains form and evolve as separate populations from silicate ones \citep{2003ARA&A..41..241D, 2001ApJ...554..778L, 2005pcim.book.....T}, suggests that ISS features  also may have} carbonaceous carriers, {although not necessarily the same carbonaceous populations as the bump carriers. In addition, the ISS features in the optical exhibiting correlations with extinction parameters in the UV (e.g., $B_3$), illustrates the result in \citet{1989ApJ...345..245C} that different parts of the curve are closely linked.} We note that these high-amplitude features are also measured with lower relative uncertainty on the amplitude, which may partially enhance the observed correlations. All these features except for ISS77 have strong positive correlations with $1/R(55)$, and thus may vary with the grain size distribution. While G25 finds that ISS77 varies with $R(55)$, we only recover a moderate ($0.3<r<0.4$) correlation. Lastly, correlations with the far-UV rise of the curve, $C_4$, can be used to suggest that carriers may originate in the molecular phase of the ISM \citep{2023ApJ...944...33V}. Here, we find little correlation with $C_4$, which means that the carriers of the ISS features may not primarily arise in the molecular regions. In addition, the weak correlations with the molecular hydrogen fraction $f(\rm H_2)$, show that the ISS features are not strictly confined to regions of molecular gas. {This is consistent with the idea that ISS features may be related to other optical extinction features such as DIBs, as suggested by M20 and M21. Since DIBs are thought to primarily originate in the diffuse atomic gas \citep{1995ARA&A..33...19H, 2006ARA&A..44..367S} {and are observed at extremely low reddening \citep{2015MNRAS.447..545B}, 
then this implies} a similar spatial origin for the carriers of ISS features.
\begin{figure*}
    \centering
    \includegraphics[width=0.7\textwidth]{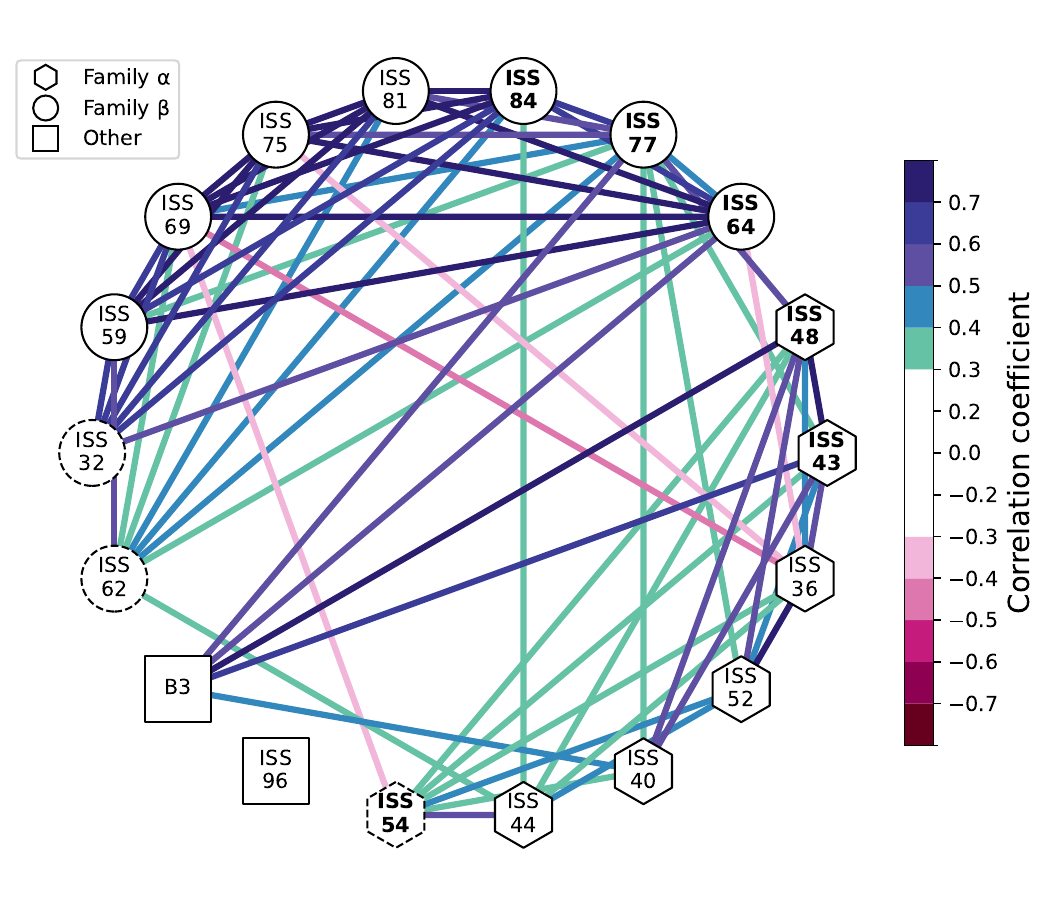}
    \caption{Circle graph of correlations between the {17} published and candidate ISS features. We present here only the moderate and strong correlations, grouping the two main ISS families. The lines represented the correlation coefficient between two ISS amplitudes, and $B_3^{A(55)}$. The nodes represent the corresponding ISS features. The hexagonal nodes represent ISS features belonging to Family $\alpha$, the circles represent those belonging to Family $\beta$. The dashed nodes represent those that have correlations with other members in the family but have a low detection significance. The 6 literature ISS features are distinguished by bold font.}
    \label{fig:corr_map_circle}
\end{figure*}

Figure~\ref{fig:corr_map_circle} presents a circle graph, in order to find trends among the ISS features, which help assess whether they may all originate from the same carrier.
The nodes represent the ISS features, and the lines between the nodes represent the correlation coefficients between pairs of features, color-coded as shown. This figure reveals that there are likely two main families of ISS features, as shown by the strong correlations between ISS48, ISS43, ISS36, ISS52, ISS40, and ISS44 as well as between ISS64, ISS77, ISS84, ISS81, ISS75, ISS69, and ISS59. We will designate the former group as Family\,$\alpha$, and the latter as Family\,$\beta$ as shown by the node shapes in the figure. While belonging to different families, ISS43 and ISS77 both show strong correlations with $B_3$ while remaining only moderately correlated with each other. This suggests that the carriers of the bump may be associated with those of each family independently.
We also find that ISS36 in Family\,$\alpha$ is moderately anti-correlated with three features in Family\,$\beta$. This further corroborates the presence of two competing grain populations, suggesting that the carriers of one family may be enhanced in environments where those of the other are suppressed.
ISS54 and ISS62 are shown to have moderate correlations with other members of Family\,$\alpha$ and $\beta$ respectively, even though they were shown to have insignificant detections. This may result from parameter covariances introduced by the simultaneous fitting procedure. Similarly, ISS32 is shown to have moderate correlations with other members of Family\,$\beta$, even though it has a weak correlation with dust column density. So, we tentatively place all three features in the respective families with the caveat of needing further investigation. Lastly, we find no strong or moderate correlations between ISS96 and the remainder of the features, so we do not place it in either family.

M21 finds that ISS77 (family $\beta$) is weaker in environments rich in molecular carbon (i.e., CO). This is consistent with the fact that we find a correlation between the amplitude of ISS77 and $B_3$ (as shown by Figure~\ref{fig:corr_map_circle}), indicating a carbonaceous origin i.e. the specific carbonaceous species responsible for ISS77 may be depleted, if carbon is preferentially locked into molecules such as CO. G25 finds correlations of ISS77 and ISS84 with $R(V)$, whereas we find moderate correlations in the present analysis. This discrepancy arises from difference in methodology, where G25 uses a method similar to Principle component analysis (PCA) with covariance in uncertainties taken into account. Thus, although our weak Pearson correlations indicate little to no pair-wise correlations between these features and $R(55)$, the correlations found in G25 may instead indicate that they may indirectly be related through a shared underlying dependence on a broader dust property rather than a direct one-to-one relationship.

\begin{figure*}
    \centering
    \includegraphics[width=\textwidth]{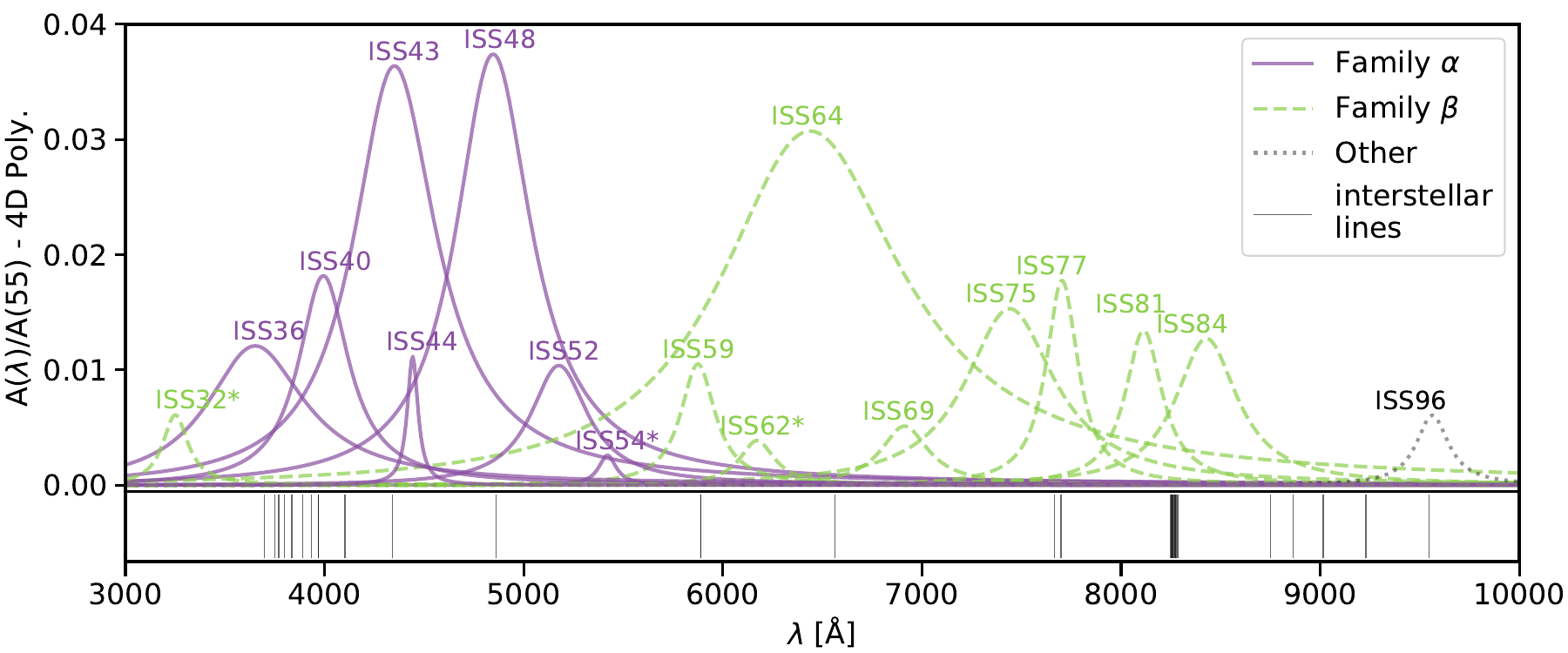}
    \caption{{\textit{Top}:} The ISS features based on their fitted parameters. The families are indicated by color and linestyle. The %dashed lines 
    features with label ``ISSnn*'' indicate features that have correlations with other members in the family but may not be dust features or have a low detection. {\textit{Bottom}: Prominent hydrogen interstellar gas lines provided for comparison.}}
    \label{fig:final_iss}
\end{figure*}
Figure\,\ref{fig:final_iss} presents a summary plot of all 17 fitted ISS features. Family\,$\alpha$ is shown in purple solid lines, Family\,$\beta$ in green dashed lines, and ISS96 identified with neither family is in dotted gray. The ISS labels marked with stars, ISS32$^*$, ISS54$^*$, and ISS62$^*$, as well as ISS96, correspond to features that either have low detection significance or are not dust features; therefore, they cannot be confidently identified as ISS features in this study. The figure shows that the Family~$\alpha$ occurs at shorter wavelengths whereas Family-$\beta$ occurs at longer wavelengths. If the two families originate from competing grain populations, as mentioned above, this wavelength separation may reflect differences in their electronic transition energies.
%Almost all the above mentioned features have little to no correlation with ISS84, ISS81, ISS75, ISS69, ISS59 and ISS62. Only ISS36, shows anti-correlations with a subset of these features, and ISS64. Additionally, we see that ISS77 also have moderate correlations with this family of ISS features. Lastly, we find that ISS64, ISS84, ISS81, ISS75, ISS69, and ISS59 are all very strongly positively correlated with each other, as evident from the dark blue triangle on the lower right of the right panel of Figure~\ref{fig:corr_map}. ISS62 is only moderately correlated with ISS77 and ISS84, strongly correlated with ISS59, and weakly correlated with the remainder of the ISS features.

\section{Summary}
\label{sec:summary}

Using HST/STIS data for 74 lines of sight that provide a statistically significant sample, we have conducted a {detailed} analysis\footnote{{All Python scripts used to conduct the work presented in this paper are publicly available on Zenodo \citep{gunasekera2026hst_iss_ext}}.} of the 6 known ISS features appearing in the literature, and the 11 newly discovered features. We follow the standard method of measuring and fitting the extinction curves outlined in \citet{2023ApJ...950...86G}. We modelled the ISS features using Drude profiles as was done in M20 to fit the wavelengths, amplitudes and widths of each feature for each line of sight in the sample. 

%Our analysis found a total of 17 ISS features. Six of these 
Of the 17 features, 6
have been previously reported in the literature--3 appearing in the original ISS paper, M20, and one each from subsequent papers, M21, Z24, and G25. We found, 5 of the 6 published ISS features, and 10 of the 11 new candidate features to have statistically significant detections well above 5$\sigma$.  ISS62 and ISS54 were calculated to have a significance of 4.3$\sigma$ and 3.2$\sigma$, respectively, in the mean extinction curve, which are statistically marginal detections. ISS54 is the weakest of the 17 features, therefore it may require much higher S/N data than available with STIS to confirm its presence. 

For all {17} features, we conduct an analysis to constrain the carriers of ISS features, as well as to confirm that the candidate features result from interstellar dust grains. 
Of the 10 statistically significant candidate features, 8 correlate with $A(55)$. However, since ISS32 also correlates with many other candidate ISS features, we retain it as an ISS candidate, giving a total of 9 ISS candidate features.
%We confirm that 9 out of the 10 statistically significant new candidate features are dust features. 
We also find that there are two families of ISS features, which we denote Family\,$\alpha$ and $\beta$. Family\,$\alpha$ includes two of the original ISS features, ISS43 and ISS48, and has strong correlations with the UV bump strength $B_3$ and $1/R(55)$. Thus Family\,$\alpha$ likely has a carbonaceous origin, and their carriers may be sensitive to dust size. 
Family\,$\beta$ includes the third original ISS feature, ISS64, as well as ISS77 both of which have moderate correlations with the bump strength, indicating a carbonaceous origin. Furthermore, all the features in Family\,$\beta$ have weak correlations with $1/R(55)$. Hence, it may be the case that the Family\,$\beta$ arises from carbonaceous carriers of a different type, without sensitivity to dust size distribution.
{Our decomposition of the extinction curve presented in Section~\ref{sec:analysis}, assumes that the six literature ISS features are well-modelled by Drude profiles. If this is not the case, some of the proposed candidate ISS features may be artifacts of the method, which illustrates the need for independent confirmation of these candidate features.}

While we have provided an in-depth study on ISS features in the Milky Way, using optical and UV data, future work is needed to probe their carriers and confirm the candidate ISS features identified here. Detections of the candidate ISS features in the laboratory or along new lines-of-sight would provide strong evidence. Additionally, correlations between the ISS features and extinction features in the mid-infrared (MIR) and near-infrared (NIR) regime would further test their potential carbonaceous nature. 
%\todo{Q3. include discussion of why ISS only in optical possible relation to DIBS, This would be analogous to the relationship between ERE and the Red Rectangle bands (RRBs), or to the infrared continuum underlying the infrared emission bands.}
The vibrational extinction features in the IR spectra can be used to probe the chemical composition of ISS carriers, while extinction measurements at longer IR wavelengths can be used to explore their size distribution. {Furthermore, since DIBs and ISS features both have not been found thus far in the UV, they are likely related, as suggested by M20 and M21. It is plausible that DIBs and ISS features may have an analogous relationship to that between the extended red emission (ERE) and the red rectangle bands (RRBs). However, future work correlating ISS features with DIBs in the optical is needed to confirm this relation, and help shed light on whether they may be two separate manifestations of similar grain populations at different size scales. Additionally, since DIBs have been found in the NIR \citep{1990Natur.346..729J} a search for ISS features in the IR may help further establish this relation.}
Future studies could also use the grouping of ISS features into families to investigate how different dust populations dominate along individual sightlines, and to explore potential correlations with physical conditions such as radiation flux and shielding, which can influence ionization, photochemistry, and dust chemistry. Finally, work is underway to extend this analysis to the Large and Small Magellanic Clouds to probe the behaviour of ISS features in different metallicity environments, which will further constrain their carriers. 

%%%%%%%%%%%%%%%%%%%%%%

%% Please use the acknowledgment and contribution environments. This will 
%% be anonomyized when the "anonymous" style option is used. 
\begin{acknowledgments}
%CG acknowledges support by HST-GO-04476, JWST-GO-02459 and HST-GO-17078.
Support for HST programs GO-04476 and GO-17078 was provided by NASA through a grant from the Space Telescope Science Institute, which is operated by the Association of Universities for Research in Astronomy, Inc., under NASA contract NAS 5-26555. Support for JWST program GO-02459 was provided by NASA through a grant from the Space Telescope Science Institute, which is operated by the Association of Universities for Research in Astronomy, Inc., under NASA contract NAS 5-03127.
MD acknowledges support from the Research Fellowship Program of the European Space Agency (ESA).

We thank Derck L. Massa, Edward L. Fitzpatrick and Catherine Zucker for their valuable discussions that helped enable this work.

{Lastly, we would like to thank the very helpful comments by the referee which significantly improved the quality of the paper.}
\end{acknowledgments}

%\begin{contribution}
%%This section gives authors the space to recognize author contributions. The text inside this environment is NOT counted towards the total word quanta. At a minimum, manuscripts are expected to include this text:

%All authors contributed equally to...

%% But authors are expected to provide more specific details, e.g. 
%%
%%SC was responsible for writing and submitting the manuscript.
%%WWM came up with the initial research concept and edited the manuscript.
%%OTS obtained the funding and edited the manuscript.
%%EBF provided the formal analysis and validation. He also edited the manuscript.
%%GEH Supervised the undergraduates, wrote the software and administers the project github and Zenodo repositories.
%%
%% Authors can use the Contributor Role Taxonomy (CRediT) at
%% for ideas on how write a good statement tailored to their needs.

%\end{contribution}

\facilities{HST(STIS)}

\software{astropy \citep{2013A&A...558A..33A,2018AJ....156..123A,2022ApJ...935..167A}, 
{matplotlib} \citep{Hunter:2007}, {numpy} \citep{numpy}, {python} \citep{python}, and {scipy} \citep{2020SciPy-NMeth}
dust\_extinction \citep{Gordon+2024:2024JOSS....9.7023G}, measure\_extinction \citep{karl_gordon_2025_15831932}}


%\appendix

%% For this sample we use BibTeX plus aasjournalv7.bst to generate the

\bibliography{ISS}{}
\bibliographystyle{aasjournalv7}



\end{document}